\documentclass[showpacs,prl,onecolumn,aps,superscriptaddress,preprintnumbers,nofootinbib]{revtex4}
\usepackage[T1]{fontenc}
\usepackage[latin9]{inputenc}
\usepackage{graphicx,hyperref}

\usepackage{amsmath,amssymb}
\usepackage{epsfig}
\usepackage{graphicx}
\usepackage{amsmath}
\usepackage{amsfonts}
\usepackage{mathrsfs}
\usepackage{pifont}
\usepackage{relsize}
\usepackage{mathtools}
\def\slashchar#1{\setbox0=\hbox{$#1$}     		
   \dimen0=\wd0                                 	
   \setbox1=\hbox{/} \dimen1=\wd1               	
   \ifdim\dimen0>\dimen1                        	
      \rlap{\hbox to \dimen0{\hfil/\hfil}}      	
      #1                                        	
   \else                                        	
      \rlap{\hbox to \dimen1{\hfil$#1$\hfil}}   	
      /                                         	
   \fi}

\renewcommand{\vec}{\boldsymbol}
\newcommand{\beq}{\begin{equation}}
\newcommand{\eeq}{\end{equation}}
\newcommand{\bea}{\begin{eqnarray}}
\newcommand{\eea}{\end{eqnarray}}
\newcommand{\baa}{\begin{array}}
\newcommand{\eaa}{\end{array}}

\def\eq#1{{Eq.~(\ref{#1})}}

\def\fig#1{{Fig.~\ref{#1}}}
\newcommand{\intl}{\int\limits}

\newcommand{\bas}{\bar{\alpha}_S}

\newcommand{\nn}{\nonumber}

\newcommand{\h}{\frac{1}{2}}

\newcommand{\Lb}{\left(}
\newcommand{\Rb}{\right)}
\def\pom{{I\!\!P}}

\newcommand{\bg}{\bar\gamma}
\newcommand{\pp}{\partial}

\renewcommand{\vec}[1]{\boldsymbol{#1}}

\newcommand{\Y}{\tilde{Y}}

\numberwithin{equation}{section}

\begin{document}
\title{Double diffraction production for dipole-dipole scattering at high energies}
\author{Carlos Contreras}

\email{carlos.contreras@usm.cl}
\affiliation{Departamento de F\'isica, Universidad T\'ecnica Federico Santa Mar\'ia,  Avenida Espa\~na 1680, Casilla 110-V, Valpara\'iso, Chile}
\author{Jos\'e Garrido}
\email{jose.garridom@sansano.usm.cl}
\affiliation{Departamento de F\'isica, Universidad T\'ecnica Federico Santa Mar\'ia,  Avenida Espa\~na 1680, Casilla 110-V, Valpara\'iso, Chile}
\affiliation{Instituto de F\'isica, Pontificia Universidad Cat\'olica de Valpara\'iso, Avenida Universidad 330, Curauma, Valpara\'iso, Chile}
\author{ Eugene Levin}
\email{leving@tauex.tau.ac.il}
\affiliation{Department of Particle Physics, School of Physics and Astronomy,
Raymond and Beverly Sackler
 Faculty of Exact Science, Tel Aviv University,\\ Tel Aviv, 69978, Israel}
\date{\today}

\begin{abstract}
 This paper is a first attempt to describe double diffractive production in dipole-dipole scattering within high energy QCD. Our approach is based on summing large BFKL Pomeron loops, which generate two clusters of hadrons separated by a rapidity gap. Unlike in the BK framework, our approach allows both clusters to carry large masses. We derive the cross section for the production of two such clusters and show that it exhibits geometric scaling. Testing this scaling behaviour against data from high-energy colliders such as the LHC will be a crucial step in establishing whether large Pomeron loops play a major role in high energy QCD dynamics, and in particular in the unitarization of the scattering amplitude.
  \end{abstract}
\maketitle

\vspace{-0.5cm}
\tableofcontents

\section{Introduction}
 
\begin{figure}[ht]
    \centering
  \leavevmode
     \includegraphics[width=14cm,height=4cm]{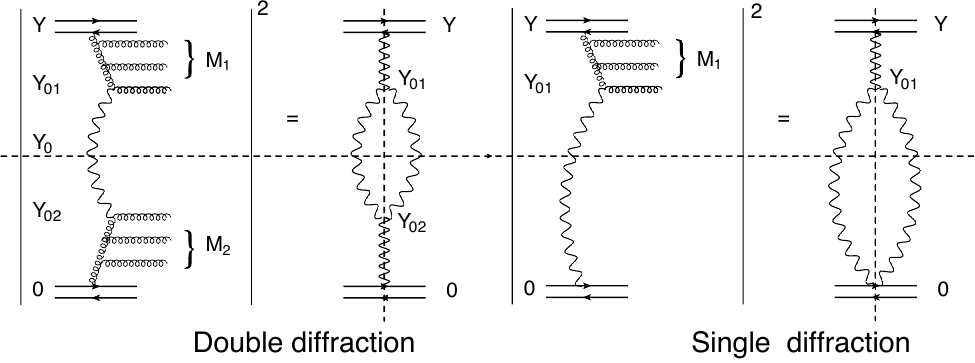}  
      \caption{The first Pomeron diagrams for double and single diffractive production. The wavy lines denote the BFKL Pomeron exchanges. The helix line denotes gluons while the arrow lines stand for heavy quarks. The produced masses are described by $\ln M^2_{1}= Y - Y_{01}\equiv Y_{m1}$ and 
   $\ln M^2_{2}= Y_{02} \equiv Y_{m2}$.  $ Y_{12} \,=\,Y_{01} \,-\,Y_{02} $ is the value of the rapidity gap in the process of double diffraction. For single diffraction, $Y_{m} \equiv Y_{m1} = \ln M^2_{1} = Y - Y_{01}$ and $Y_{01}$ denote the produced mass and the rapidity gap, respectively. The two cascades in \eq{MPSIDD} interact at rapidity $Y=Y_0$ (see \fig{mpsi} and \fig{mpsidd}).}
\label{dodif}
   \end{figure}

In this paper we are going to discuss the process of double diffractive production. In \fig{dodif} one can see the first Pomeron diagram for this process, in which we put all kinematic notations.
We consider the scattering of two mesons, each consisting of two heavy quarks, so that the perturbative QCD approach is applicable. We are interested in the production of two systems of hadrons with large masses $M_1$ and $M_2$. In this case double diffractive production can be described by the exchange of BFKL Pomerons \cite{BFKL}\footnote{BFKL stands for Balitsky, Fadin, Kuraev and Lipatov.}, as shown in the second diagram of \fig{dodif}.

The equations for single diffraction have been derived in Refs.~\cite{KOLE,KOLEB}, building on the dipole approach to high energy QCD \cite{MUDI}, which reproduces the nonlinear evolution equation for the scattering of a dilute parton system off a dense one \cite{GLR,MUQI,B,K,JIMWLK1,JIMWLK2}. However, despite considerable progress in the theoretical understanding of high energy QCD \cite{BFKL,KOLEB,MUSA,LETU,LELU,LIP,KO1,LE11,RS,KLremark2,
SHXI,KOLEV,nestor,LEPRI,LMM,LEM,IAMU,MUT,MUPE,IIML,
LIFT,GLR,GLR1,MUQI,MUDI,NAPE,BART,BJKP,GGM,MV, KOLE,BRN,BRAUN,B,K,KOLU,JIMWLK1,JIMWLK2,utmp,AKLL,KOWI, KOLU11,KOLUD,BA05,SMITH,KLW,KLLL1,KLLL2,LEPR,LE1,LEDIDI,
LEDIA,LEAA,LELT}, the evolution equations for dilute-dilute and dense-dense parton scattering remain unknown. The central unsolved theoretical problem is the summation of the Pomeron loops, of which \fig{dodif} shows one example; we therefore need a way to handle these loops.

\begin{figure}[ht]
    \centering
  \leavevmode
     \includegraphics[width=5cm]{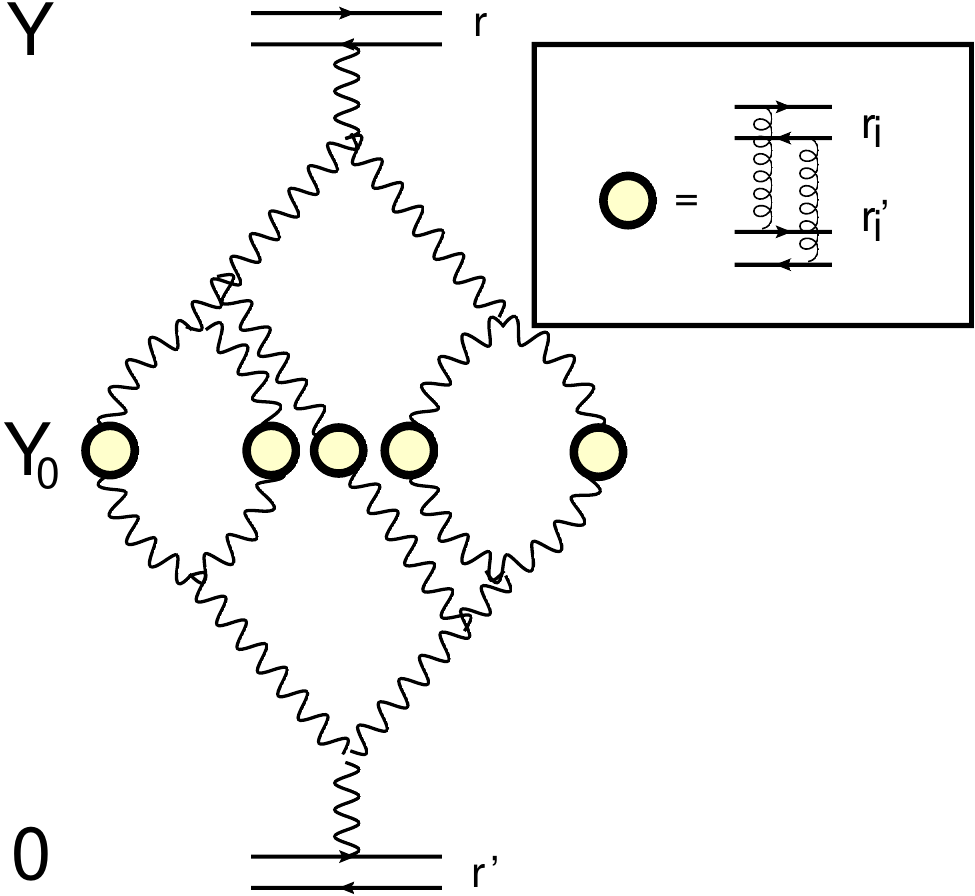}  
      \caption{ Summing large Pomeron loops. The wavy lines denote the BFKL Pomeron exchanges. The circles denote the amplitude $\gamma$.}
\label{mpsi}
   \end{figure}

We sum the large Pomeron loops using $t$-channel unitarity, recast in a form convenient for the dipole approach to the CGC in Refs.~\cite{MUSA,IAMU,KOLEB,MUDI,LELU,KO1,LE11} (see \fig{mpsi}).
The analytic expression for the imaginary part of the scattering amplitude then takes the form \cite{LELU,KO1,LE1}:  
\bea \label{MPSI}
     && N\Lb Y, \vec{r},\vec{r}';  \vec{b}\Rb\,=\\
     &&\,\sum^\infty_{n=1}\,\Lb -1\Rb^{n+1}\,n!\int  \prod^n_{i=1} d^2 r_i\,d^2\,r'_i\,d^2 b'_i 
     \int \!\!d^2 \delta b_i\, \gamma^{BA}\Lb \vec{r}_i,\vec{r}'_i, \vec{b}_i -  \vec{b}'_i\equiv \delta \vec{b} _i\Rb 
    \,\rho_n\Lb Y - Y_0,\vec{r},\vec{b}; \{ \vec{r}_i,\vec{b}_i\}\Rb\,\rho_n\Lb Y_0,\vec{r}',\vec{b}'; \{ \vec{r}'_i,\vec{b}'_i\}\Rb \nn
\eea
      
$\gamma^{BA}$ is the scattering amplitude of two dipoles in the Born approximation of perturbative QCD. The dipole densities $\rho_n(Y, \vec{r},\vec{b};\{ \vec{r}_i,\vec{b}_i\})$ have been introduced in Ref.~\cite{LELU} as follows:
\beq \label{PD}
\rho_n\Lb Y\,-\,Y_0,\vec{r},\vec{b}; \{ \vec{r}_i,\vec{b}_i\}\Rb\,=\,\frac{1}{n!}\,\prod^n_{i =1}
\,\frac{\delta}{\delta
u_i } \,Z\left(Y\,-\,Y_0, \vec{r},\vec{b};\,[u_i] \right)\Big{|}_{u_i=1}
\eeq
where  the generating functional $Z$ is
\beq \label{ZGEN}
Z\Lb Y, \vec{r},\vec{b}; [u_i]\Rb\,=\,\sum^{\infty}_{n=1}\int P_n\Lb Y,\vec{r},\vec{b};\{\vec{r}_i,\vec{b}_i\}\Rb \prod^{n}_{i=1} u\Lb \vec{r}_i,\vec{b}_i\Rb\,d^2 r_i\,d^2 b_i
\eeq
where $u\Lb \vec{r}_i,\vec{b}_i\Rb \equiv\,u_i$ is an arbitrary function and $P_n$ is the probability of finding $n$ dipoles with the given kinematics.
The initial and  boundary conditions for the BFKL cascade which stems from one dipole have the following form for the functional $Z$:
 
\begin{subequations}
\bea
Z\Lb Y=0, \vec{r},\vec{b}; [u_i]\Rb \,\,&=&\,\,u\Lb \vec{r},\vec{b}\Rb;~~~~~~~~Z\Lb Y, \vec{r},\vec{b};[u_i=1]\Rb \,\,=\,\, 1;\label{ZIC}\\
\rho_1\Lb Y=0, \vec{r},\vec{b}; \vec{r}_1,\vec{b}_1\Rb\,\,&=&\,\,\delta^{(2)}\Lb \vec{r} - \vec{r}_1\Rb \delta^{(2)}\Lb \vec{b} - \vec{b}_1\Rb ;~~~~~\rho_n\Lb 
 Y=0, \vec{r},\vec{b}; \{\vec{r}_i, \vec{b}_i\}\Rb \,\,=\,\,0 ~\, \mbox{at}~\,n\geq 2;\label{ZSR}
\eea
\end{subequations}
  In \eq{MPSI} $\vec{b}_i\,=\,\vec{b} \,-\,\vec{b}_i'$.

In Ref.~\cite{LELT} we found $ \rho_n\Lb Y - Y_0, \vec{r},\vec{b}; \{ \vec{r}_i,\vec{b}_i\}\Rb$, which in the saturation region take the form:
\beq \label{RHONF} 
   \rho_n\Lb Y - Y_0, \vec{r},\vec{b};\{ \vec{r}_i,\vec{b}_i\}\Rb\,\,=\,\,C_n  \prod_{i=1}^n G_{\pom} \Lb z_i\Rb~~~~\mbox{with}~~~~z_i \,= \,\kappa\,\bas\,\Lb Y - Y_0\Rb \,+\,\xi_{\vec{r}_i, \vec{r}}
   \eeq
   where $r$ is the size of the incoming dipole. The scaling variable $z_i$ is equal to
   $z_i \,=\,\ln\Lb  r^2_i\,Q^2_s\Lb Y, \vec{b}\Rb\Rb$, where $Q_s$ is a new dimensional scale: the saturation momentum~\cite{GLR} which has the following $Y$ dependence~\cite{GLR,MUT,MUPE}:
 \beq \label{QS}
 Q^2_s\Lb Y, \vec{b}\Rb\,\,=\,\,Q^2_s\Lb Y=0, \vec{b}\Rb \,e^{\bas\,\kappa \,Y\,-\,\,\frac{3}{2\,\bg} \ln \Lb \bas\,Y\Rb }
 \eeq 
 \eq{RHONF} shows geometric scaling behaviour \cite{GS,IIML} as a function of the single variable $z_i$. The Green's function of the BFKL Pomeron in this region is known~\cite{MUT}:
 \beq \label{POMZ}
 G_{\pom} \Lb z_i\Rb\,\,=\,\,e^{\bar{\gamma}\,z_i}
 \eeq 
   The value of $\kappa$ and $\bar{\gamma}$ are  determined by the BFKL kernel (see for example Ref.~\cite{KOLEB}).
   
   However, in this paper we will not use \eq{RHONF}; instead we will find $\rho_n$ in \eq{MPSI} using the equation for single diffraction and the simple expression for the densities $\rho^A_n$ for a nucleus with radius $r'=R_A = A^{1/3} R_N$ \footnote{$R_N$ is the radius of a nucleon.}, assuming that the nucleus is a bag of dipoles with size $r'$ that do not interact with each other at $Y_0  = 0$. 
It has the form
\beq \label{DD41}
\rho^A_n\Lb Y_0=0, \vec{r}', \vec{b}';  \{ \vec{r}'_i,\vec{b}'_i\}\Rb \,\,=\,\,\frac{S^n_A\Lb b'\Rb}{n!}\prod^n_{i=1} \delta^{(2)}\Lb \vec{r}' - \vec{r}'_i\Rb \delta^{(2)}\Lb \vec{b}' - \vec{b}_i\Rb ;
\eeq
 The factor $S^n_A\Lb b' \Rb$ gives the probability of finding $n$ nucleons (dipoles) in a nucleus at impact parameter $b'$. Finally,
  $G_{\pom} \Lb z_i\Rb  $ is the solution of the BFKL equation (the Green's function of the BFKL Pomeron) in the saturation region, where it exhibits
   geometric scaling behaviour~\cite{GS,IIML}. The BFKL equation takes the following general form:
\bea \label{BFKL}
  &&\frac{\partial\,G_{\pom}\Lb Y,  \vec{x}_{01}, \vec{r}_i,\vec{b} \Rb }{\partial\,Y} \,\,=\,\, \frac{\bas}{2\,\pi} \intl d^2 x_2 \frac{x_{01}^2}{x_{02}^2\,x_{12}^2}\,\nn\\ 
  &&\times\,\left[ G_{\pom}\Lb Y,   \vec{x}_{12}, \vec{r}_i,\vec{b} \,-\, \frac{\vec{x}_{02}}{2} \Rb\,\,+\,\,G_{\pom}\Lb Y,  \vec{x}_{02}, \vec{r}_i,\vec{b} \,-\, \frac{\vec{x}_{12}}{2} \Rb \,\,-\,\,G_{\pom}\Lb Y,  \vec{x}_{01}, \vec{r}_i, \vec{b} \Rb\right]
\eea
  with the initial condition which corresponds to the exchange of two gluons in $t$-channel (see Appendix~\ref{appendix:A} for its solution).

\section{Summing large Pomeron loops: diffraction in the dipole-dipole scattering}
\label{sumLPL}

  
\subsection{General formulae}
\label{sumLPLs1}

  ~
  
  The equation for the single diffraction in dipole-nucleus collisions has the following form in QCD \cite{KOLE,KOLEB}:
  \beq \label{KL}
  \frac{\partial  S^D\Lb \vec{x}_0,\vec{x}_1, Y,Y_{01}\Rb}{\partial Y}\,\,=\,\,
  \frac{\bas}{2\,\pi}\int d^2\,x_{2}\, \frac{ x^2_{01}}{x^2_{02}\,x^2_{12}}\,\Big\{  
  S^D\Lb \vec{x}_1,\vec{x}_2, Y,Y_{01}\Rb  S^D\Lb \vec{x}_0,\vec{x}_2, Y,Y_{01}\Rb  \,\,-\,\,S^D\Lb \vec{x}_0,\vec{x}_1, Y,Y_{01}\Rb  \Big\}
  \eeq
  
  where 
  \beq \label{DD11}
  S^D\Lb \vec{x}_0,\vec{x}_1, Y,Y_{01}\Rb\,\,=\,\,1\,- 2\,N \Lb \vec{x}_0,\vec{x}_1, Y\Rb \,+\,N^D\Lb \vec{x}_0,\vec{x}_1, Y,Y_{01}\Rb 
  \eeq
where $N$ is the amplitude for the elastic scattering and $N^D$ is the single diffraction cross section with the rapidity gap $ \geq \,Y_0$. All vectors are two dimensional. 

The Balitsky--Kovchegov (BK) nonlinear equation~\cite{B,K} for the scattering amplitude $N \Lb \vec{x}_0,\vec{x}_1, Y\Rb$  has the same form as \eq{KL} but for
\beq\label{DD12}
S\Lb\vec{x}_0,\vec{x}_1, Y\Rb \,\,=\,\,1\,-\,N \Lb \vec{x}_0,\vec{x}_1, Y\Rb
\eeq

The difference between them is only in the initial condition, which for \eq{KL} has the form:
\beq \label{DD13}
S^D\Lb \vec{x}_0,\vec{x}_1, Y=Y_{01},Y_{01}\Rb\,\,=\,\,S^2\Lb  \vec{x}_0,\vec{x}_1, Y_{01}\Rb
\eeq

To calculate the diffraction cross sections in QCD, we need to modify \eq{MPSI} by (i) introducing two scattering amplitudes: $\gamma$ for the scattering amplitude in \fig{dodif}
 and $\bar{\gamma}$ for the complex conjugate amplitude (see
 Ref.~\cite{LEPRI} for details); and (ii) setting a new definition for parton densities $\rho^{D}_{n, k}(Y\,-\,Y_0,Y_{01},\vec{r},\vec{b};\vec{r}_1, \vec{b}_1,\,
\ldots\,,\vec{r}_n, \vec{b}_n; \ldots\,,\vec{r}_k, \vec{b}_k)$:

\beq \label{PDD}
\rho^{D}_{n, k}(Y\,-\,Y_0,Y_{01},\vec{r},\vec{b};\{ \vec{r}_i,\vec{b}_i\},\{ \vec{r}_j,\vec{b}_j\})\,=\,\frac{1}{n!}\,\prod^n_{i =1}
\,\frac{\delta}{\delta
u_i } \,  \frac{1}{k!}\,\prod^k_{j=1}
\,\frac{\delta}{\delta
\bar{u_j} }Z^D\left(Y\,-\,Y_0, Y_{01},\vec{r},\vec{b};\,[u_i] ,\,[\bar u_j]\right)\Big{|}_{u_i=1,\bar u_j= 1}
\eeq
with $Z^D\left(Y\,-\,Y_0,Y_{01},\vec{r},\vec{b};\,[u_i] ,\,[\bar u_j]\right) = S ^D\Lb Y\,-\,Y_0,Y_{01},\vec{r},\vec{b}; \gamma = 1 - u_i, \bar{\gamma} = 1 - \bar u_j\Rb$.

\begin{figure}[ht]
    \centering
  \leavevmode
     \includegraphics[width=10cm]{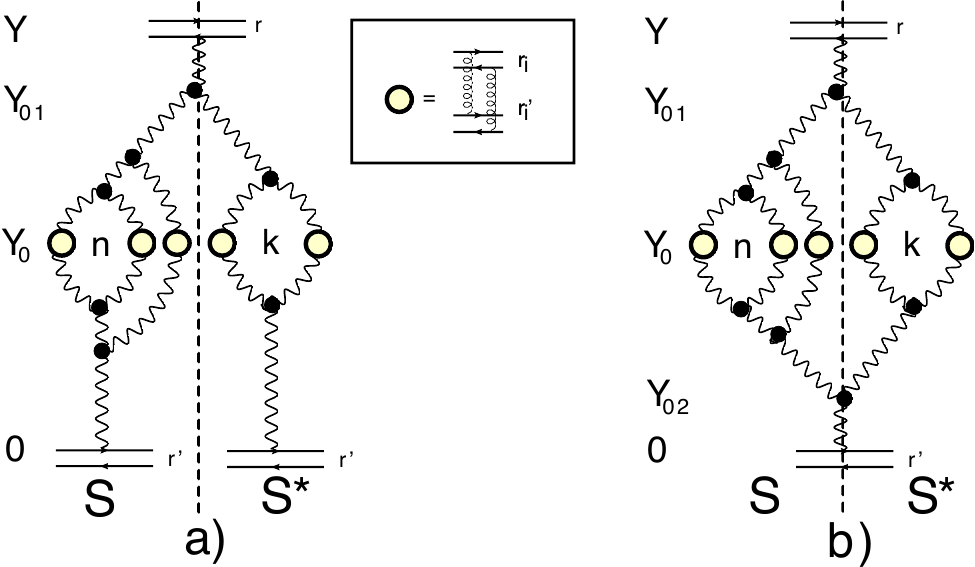}  
      \caption{ Summing  large Pomeron loops for diffraction in the dipole-dipole scattering.
The wavy lines denote the  BFKL Pomeron exchanges.  The yellow circles denote the amplitude $\gamma$. The black circle denotes the triple Pomeron vertices. \fig{mpsidd}-a shows the single diffraction cross section. For simplicity we put only the BFKL Pomeron exchange for the production of the mass $\ln M^2_{1}= Y - Y_{01}$. \fig{mpsidd}-b describes the double diffraction. Only  the BFKL Pomeron exchanges are shown for production $\ln M^2_{1} = Y - Y_{01}$ and $\ln M^2_{2}= Y_{02}- 0$. $S$ and $S^*$ stand for the $S$-matrix and its conjugate.}
\label{mpsidd}
   \end{figure}

Using these densities we can rewrite \eq{MPSI} in the following form for single diffraction in the dipole-dipole scattering (see \fig{mpsidd}-a):
 \bea \label{MPSID}
     && S^D_{dd}\Lb Y-Y_{01},Y_{0},\vec{r},\vec{r}' ;  \vec{b}\Rb\,=\\
     &&\,\sum^\infty_{n=0}\,\Lb -1\Rb^{n}\,n!\,\sum^\infty_{k=0}\,\Lb -1\Rb^{k}\,k!\int  \prod^n_{i=1} d^2 r_i\,d^2\,r'_i\,d^2 b'_i \prod^k_{j=1} d^2 r_j\,d^2\,r'_j\,d^2 b'_j\, \rho^{D}_{n, k}(Y\,-\,Y_0,Y_{01},\vec{r},\vec{b};\{ \vec{r}_i,\vec{b}_i\},\{ \vec{r}_j,\vec{b}_j\})  \nn\\
&&\times    
     \int \!\!d^2 \delta b_i\, \gamma^{BA}\Lb \vec{r}_i,\vec{r}'_i, \vec{b}_i -  \vec{b}'_i\equiv \delta \vec{b} _i\Rb  \int \!\!d^2 \delta b_j\,\bar{ \gamma}^{BA}\Lb \vec{r}_j,\vec{r}'_j, \vec{b}_j-  \vec{b}'_j\equiv \delta \vec{b} _j\Rb     \,\rho_n\Lb Y_{0},\vec{r},\vec{b}; \{ \vec{r}'_i,\vec{b}'_i\}\Rb  \rho_k\Lb  Y_{0},\vec{r},\vec{b}; \{ \vec{r}'_j,\vec{b}'_j\}\Rb  \nn
      \eea

For double diffraction we have:
 \bea \label{MPSIDD}
     && S^{DD}\Lb Y-Y_{01},Y_{02},Y_{01}-Y_{02},\vec{r},\vec{r}' ;  \vec{b}\Rb\,=\\
     &&\,\sum^\infty_{n=0}\,\Lb -1\Rb^{n}\,n!\,\sum^\infty_{k=0}\,\Lb -1\Rb^{k}\,k!\int  \prod^n_{i=1} d^2 r_i\,d^2\,r'_i\,d^2 b'_i \prod^k_{j=1} d^2 r_j\,d^2\,r'_j\,d^2 b'_j \rho^{D}_{n, k}(Y\,-\,Y_0,Y_{01},\vec{r},\vec{b};\{ \vec{r}_i,\vec{b}_i\},\{ \vec{r}_j,\vec{b}_j\})   \nn\\
&&\times    
     \int \!\!d^2 \delta b_i\, \gamma^{BA}\Lb \vec{r}_i,\vec{r}'_i, \vec{b}_i -  \vec{b}'_i\equiv \delta \vec{b} _i\Rb  \int \!\!d^2 \delta b_j\, \bar{\gamma}^{BA}\Lb \vec{r}_j,\vec{r}'_j, \vec{b}_j-  \vec{b}'_j\equiv \delta \vec{b} _j\Rb  \rho^{D}_{n, k}(Y_0,Y_{02},\vec{r}',\vec{b}';\{ \vec{r}'_i,\vec{b}'_i\},\{ \vec{r}'_j,\vec{b}'_j\})    \,\,\, \nn
      \eea

To carry out calculations using \eq{MPSID} and \eq{MPSIDD}, we note that
$t$-channel unitarity for the BFKL Pomeron in QCD takes the following form \cite{MUSA} (see notations in \eq{MPSI}):
\beq \label{UT}
G_{\pom} \Lb Y, \vec{r},\vec{r}',\vec{b}\Rb \,\,=\,\,\intl d^2 r_i \,d^2 b_i\, d^2r'_i\, d^2 b'_i\, 
G_{\pom}\Lb Y - Y_0, \vec{r}, \vec{b} - \vec{b}_i, \vec{r}_i\Rb \gamma^{BA}\Lb \vec{r}_i,\vec{r}'_i,\delta \vec{b_i}\Rb\,G_{\pom}\Lb  Y_0; \vec{r}', \vec{b}'_i, \vec{r}'_i\Rb
\eeq
where $\gamma^{BA}\Lb \vec{r}_i, \vec{r}'_i,\delta \vec{b_i}\Rb$ is the scattering amplitude of two dipoles in Born approximation of perturbative QCD.

\subsection{Equation for double diffraction} 

In this section we attempt to write the evolution equation for double diffraction production in dipole-dipole scattering, based on summing large Pomeron loops. We use the observation that in \eq{MPSI}, for the dipole-dipole scattering amplitude, the structure of the dipole cascade is the same as in BK evolution, both for $Y>Y_0$ and for $Y<Y_0$. Therefore, for the fastest and the slowest dipoles we can trust the BK equations.

With this in mind, for $Y -Y_0  >Y_{m1}> 0$ we can write the following equation for double diffraction\footnote{All notations are in \fig{dodif}.} :
\bea \label{KLDD}
&& \frac{\partial S^{DD}\Lb \vec{x}_0,\vec{x}_1, \vec{x}'_0,\vec{x}'_1,Y_{m1},Y_{12},Y_{m2}\Rb}{\partial Y_{m1}}\,=\,
\frac{\bas}{2\,\pi}\int d^2x_2\,\frac{x^2_{01}}{x^2_{02}\,x^2_{12}}\\
&&\times\,\Big\{
S^{DD}\Lb \vec{x}_1,\vec{x}_2, \vec{x}'_0,\vec{x}'_1,Y_{m1},Y_{12},Y_{m2}\Rb\,S^{DD}\Lb \vec{x}_0,\vec{x}_2,  \vec{x}'_0,\vec{x}'_1,Y_{m1},Y_{12},Y_{m2} \Rb
-S^{DD}\Lb \vec{x}_0,\vec{x}_1, \vec{x}'_0,\vec{x}'_1,Y_{m1},Y_{12},Y_{m2}\Rb\Big\}, \nn
\eea
with the initial condition
\beq \label{KLDD1}
 S^{DD}\Lb \vec{x}_0,\vec{x}_1, \vec{x}'_0,\vec{x}'_1,Y_{m1}=0,Y_{12},Y_{m2}\Rb\,=\,
 S^{D}_{dd}\Lb \vec{x}_0,\vec{x}_1, \vec{x}'_0,\vec{x}'_1,Y,Y_{m2}\Rb \eeq
where $S^{D}_{dd}$ describes single diffraction in dipole-dipole scattering.

For $Y_{m2} < Y_0$ the same equation takes the form
\bea \label{KLDD2}
&& \frac{\partial S^{DD}\Lb \vec{x}_0,\vec{x}_1, \vec{x}'_0,\vec{x}'_1,Y_{m1},Y_{12},Y_{m2}\Rb}{\partial Y_{m2}}\,=\,-
\frac{\bas}{2\,\pi}\int d^2x'_2\,\frac{x'^2_{01}}{x'^2_{02}\,x'^2_{12}}\\
&&\times\,\Big\{
S^{DD}\Lb \vec{x}_0,\vec{x}_1, \vec{x}'_1,\vec{x}'_2,Y_{m1},Y_{12},Y_{m2}\Rb\,S^{DD}\Lb \vec{x}_0,\vec{x}_1,  \vec{x}'_0,\vec{x}'_2,Y_{m1},Y_{12},Y_{m2} \Rb
-S^{DD}\Lb \vec{x}_0,\vec{x}_1, \vec{x}'_0,\vec{x}'_1,Y_{m1},Y_{12},Y_{m2}\Rb\Big\}, \nn
\eea
and the initial condition of \eq{KLDD1} becomes
\beq \label{KLDD3}
 S^{DD}\Lb \vec{x}_0,\vec{x}_1, \vec{x}'_0,\vec{x}'_1,Y_{m1},Y_{12},Y_{m2}=0\Rb\,=\,
 S^{D}_{dd}\Lb \vec{x}_0,\vec{x}_1, \vec{x}'_0,\vec{x}'_1,Y_{m1},Y\Rb\eeq

These equations must be supplemented by the equations for single diffraction in dipole-dipole scattering, which have a similar form:
\begin{subequations}
\bea
&&\frac{\partial S^{D}_{dd}\Lb \vec{x}_0,\vec{x}_1, \vec{x}'_0,\vec{x}'_1,Y_{m1},Y_{12}\Rb}{\partial Y_{m1}}\,=\,\frac{\bas}{2\,\pi}\int d^2x_2\,\frac{x^2_{01}}{x^2_{02}\,x^2_{12}} \label{KLDD4}\\
&&\times\,\Big\{
S^{D}_{dd}\Lb \vec{x}_1,\vec{x}_2, \vec{x}'_0,\vec{x}'_1,Y_{m1},Y_{12}\Rb\,S^{D}_{dd}\Lb \vec{x}_0,\vec{x}_2, \vec{x}'_0,\vec{x}'_1,Y_{m1},Y_{12}\Rb
-S^{D}_{dd}\Lb \vec{x}_0,\vec{x}_1, \vec{x}'_0,\vec{x}'_1,Y_{m1},Y_{12}\Rb\Big\},\nn\\
&&-\frac{\partial S^{D}_{dd}\Lb \vec{x}_0,\vec{x}_1, \vec{x}'_0,\vec{x}'_1,Y_{12},Y_{m2}\Rb}{\partial Y_{m2}}\,=\,-
\frac{\bas}{2\,\pi}\int d^2x'_2\,\frac{x'^2_{01}}{x'^2_{02}\,x'^2_{12}} \label{KLDD5}\\
&&\times\,\Big\{
S^{D}_{dd}\Lb \vec{x}_0,\vec{x}_1,  \vec{x}'_1,\vec{x}'_2,Y_{12} ,Y_{m2}\Rb\,S^{D}_{dd}\Lb \vec{x}_0,\vec{x}_1, \vec{x}'_0,\vec{x}'_2,Y_{12},Y_{m2}\Rb
-S^{D}_{dd}\Lb \vec{x}_0,\vec{x}_1, \vec{x}'_0,\vec{x}'_1,Y_{12},Y_{m2}\Rb\Big\}. \nn
\eea
\end{subequations}
The initial conditions for both equations have the form
\beq \label{KLDD6}
S^{D}_{dd}\Lb \vec{x}_0,\vec{x}_1,\vec{x}'_0,\vec{x}'_1,Y,Y_{m2}=0\Rb\,=\,S^{D}_{dd}\Lb \vec{x}_0,\vec{x}_1,\vec{x}'_0,\vec{x}'_1,Y_{m1}=0,Y\Rb\,=\,\Lb S^d_d\Lb \vec{x}_0,\vec{x}_1,\vec{x}'_0,\vec{x}'_1,Y\Rb\Rb^2,
\eeq
where $S^d_d$ is the elastic $S$-matrix for dipole-dipole scattering, discussed in Appendix~\ref{appendix:B}. 

In the rest of the paper we are going to discuss more detailed solution based of the direct application of \eq{MPSIDD}.

\section{ Double diffraction in the one-dimensional model}
\label{DDIF1d}

\subsection{ Single diffraction} 


In the one-dimensional model~\cite{ACJ,AAJ,JEN,ABMC,CLR,CIAF,MUSA,nestor,RS,SHXI,KOLEV,BMMSX,BIT,LEPRI,utm,utmm,utmp,LETMP} we neglect the dependence of the amplitude with respect to the transverse coordinate and \eq{KL} takes the form:
 \beq \label{KL1}
  \frac{\partial  S^D\Lb Y,Y_{01}\Rb}{\partial Y}\,\,=\,\,
  \Delta_{\pom}\,\Big\{  
  S^D\Lb Y,Y_{01}\Rb^2  \,\,-\,\,S^D\Lb Y,Y_{01}\Rb  \Big\}
  \eeq
where $\Delta_{\pom} $ is the intercept of the BFKL Pomeron. Introducing a new variable $\Y = \Delta_{\pom}\,Y$ one can see that \eq{KL1} has the general solution:

\beq \label{DD14}
S^D\Lb \Y,\Y_{01}\Rb \,\,=\,\,\frac{1}{1\,\,+\,\,\Phi\Lb \Y_{01}\Rb e^{ \Y - \Y_{01}}}
\eeq
with the equation for $\Phi$:

\beq \label{DD15}
\frac{1}{1 + \Phi\Lb \Y_{01}\Rb} \,\,=\,\, S^2\Lb \Y_{01}\Rb
\eeq
Solution for $S$ of the same equation is

\beq \label{DD16}
S\Lb \gamma, \Y\Rb \,\,=\,\,\frac{1 - \gamma}{1\,\,+\,\,\gamma\Lb e^{ \Y } \,-\,1\Rb}\,\,=\,\,\frac{1 }{1\,\,+\,\,\gamma_R \,e^{ \Y } }
\eeq
with the initial condition $S\Lb \gamma, \Y=0\Rb\,=\,1 - \gamma$ where $\gamma$ is the amplitude of the interaction with the target at low energies, $\gamma_R = \gamma/(1 - \gamma)$. Plugging this equation into \eq{DD15} we obtain:
\beq \label{DD17}
S^D\Lb \Y,\Y_{01}\Rb \,\,=\,\,\frac{1}{1\,\,-\,\, e^{ \Y _{m1}}\,\,+\,\, e^{ \Y_{m1}}\ S^{-2}\Lb \gamma,\Y_{01}\Rb}=\frac{1}{1+  \gamma_R\,e^{\Y}\Lb 2  +\gamma_R
e^{\Y_{01}}\Rb}\eeq
The cross section for the diffractive production with fixed rapidity gap $Y_{01}$ is equal to \cite{KOLE}
\beq\label{SD1}
\sigma^{D}(\Y,\Y_{01}) \,\,=\,\, - \,\frac{\pp N^D(\Y,\Y_{01})}{\pp \Y_{01}} \,=\, -\frac{\pp S^D(\Y,\Y_{01})}{\pp \Y_{01}}\eeq
In \eq{SD1} we use \eq{DD11} calculating $\frac{\pp N^D(\Y,\Y_{01})}{\pp \Y_{01}}$.
For \eq{DD17} it takes the form\footnote{We neglect the contribution of the elastic scattering in the process of diffractive production.}
\beq\label{SD2}
\sigma^{D}(\Y,\Y_{01}) \,\,=\,\,\frac{ \gamma^2_R e^{\Y+\Y_{01}} }{\Lb 
1\,+\,\gamma_R e^{\Y}\Lb 2  +\gamma_R
e^{\Y_{01}}\Rb\Rb^2}
\eeq
where $\Y = \Y_{m1} + \Y_{01}$.

\begin{figure}[ht]
    \centering
  \leavevmode
     \includegraphics[width=9cm]{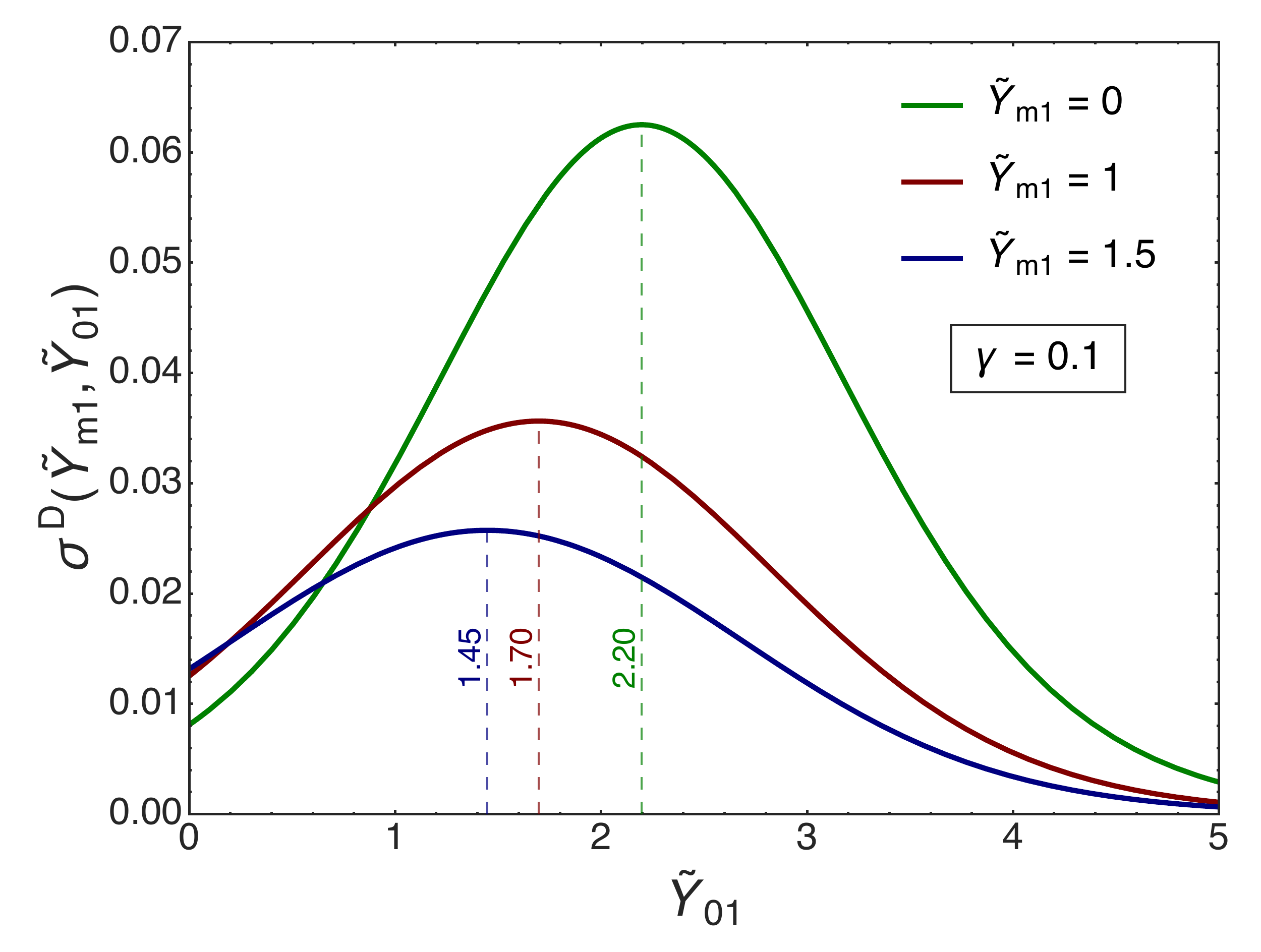}  
      \caption{$\sigma^D$ versus $\Y_{01} $ at fixed $\Y_{m1}$. $\gamma =0.1$.}
\label{xsdif}
   \end{figure}

We plot this cross section in \fig{xsdif} for $\gamma=0.1$; one can see that $\sigma^D$ decreases as a function of $\Y_{01}$ and also decreases with increasing $\Y_{m1}$. The characteristic behaviour exhibits a maximum at $Y_0 =Y_{max}$, and $Y_{max}$ decreases as $\Y_{m1}$ increases. All these features have been discussed in Ref.~\cite{KOLE}. Coming back to \eq{DD17} one can see that this solution can be rewritten in the following form:
\beq \label{DD18}
S^D\Lb \Y,\Y_{01}\Rb \,\,=\,\frac{1}{1\,+\, \gamma_R\,e^{\Y_{m1} + \Delta z\Lb \Y_{01}\Rb}}\,\equiv \,S\Lb \Y_{m1} + \Delta z\Lb \Y_{01}\Rb\Rb~\mbox{with}~ \Delta z\Lb \Y_{01}\Rb = \Y_{01}\,+\,\ln\Lb 2 + \gamma_R e^{\Y_{01}}\Rb;
\eeq
The function $ \Delta z\Lb \Y_{01}\Rb $ can be found from the following equation:
\beq \label{DD19}
S\Lb  \Delta z\Lb \Y_{01}\Rb\Rb \,=\, S^2\Lb \Y_{01}\Rb
\eeq

This is easy to understand: the equation for $S^D$ has the same form as the Balitsky--Kovchegov equation for the elastic scattering matrix $S$.  The two therefore have the same general solution (see \eq{DD18}) and differ only in the initial condition (see \eq{DD19}).

\subsection{Double diffraction} 

\begin{figure}[ht]
    \centering
  \leavevmode
     \includegraphics[width=13cm]{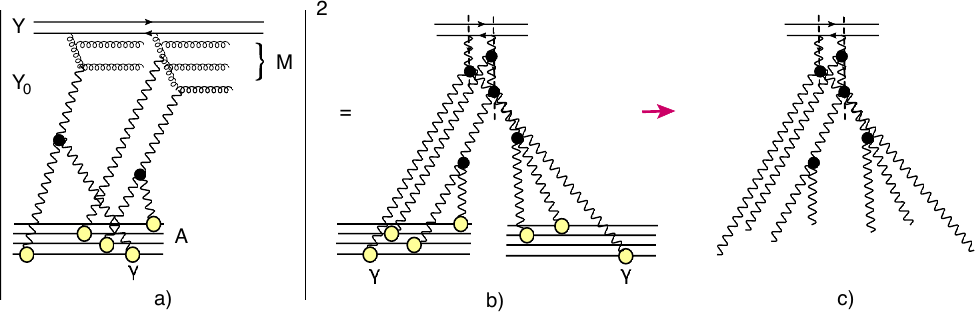}  
      \caption{\fig{sdpom}-a and b: An example of the Pomeron diagrams that contribute to the single diffractive production. Wavy lines are Pomerons. The vertical dashed lines show the cut Pomerons. The black circle denotes the triple Pomeron vertices while the yellow circle shows the scattering amplitude with the nucleon. \fig{sdpom}-c shows the contribution of these diagrams to \eq{MPSI}. }
\label{sdpom}
   \end{figure}

It is well known that high energy QCD can be formulated as the BFKL Pomeron calculus: the exchange and interactions between BFKL Pomerons~\cite{GLR,MUQI,KOLE,BART,BRN,BRAUN,AKLL,KOLU11,KLW,KLLL2}. In particular, in Ref.~\cite{LELT} it is shown that $\rho_n$ in \eq{MPSI} can be found in this approach. In \fig{sdpom} we give an example of the Pomeron diagram that describes the single diffractive production for the interaction of the dipole with a nucleus. Our main idea is to use \eq{DD41} for
$\rho^A_n\Lb Y_0=0, \vec{r}', \vec{b}';  \{ \vec{r}'_i,\vec{b}'_i\}\Rb$ and \eq{MPSI} and find the Pomeron structure of $\rho_n$ of the fast dipole  for the diffractive production. An example of such Pomeron diagrams can be seen in \fig{sdpom}-c.

\begin{figure}[ht]
    \centering
  \leavevmode
  \begin{tabular}{c c c}
     \includegraphics[width=5cm]{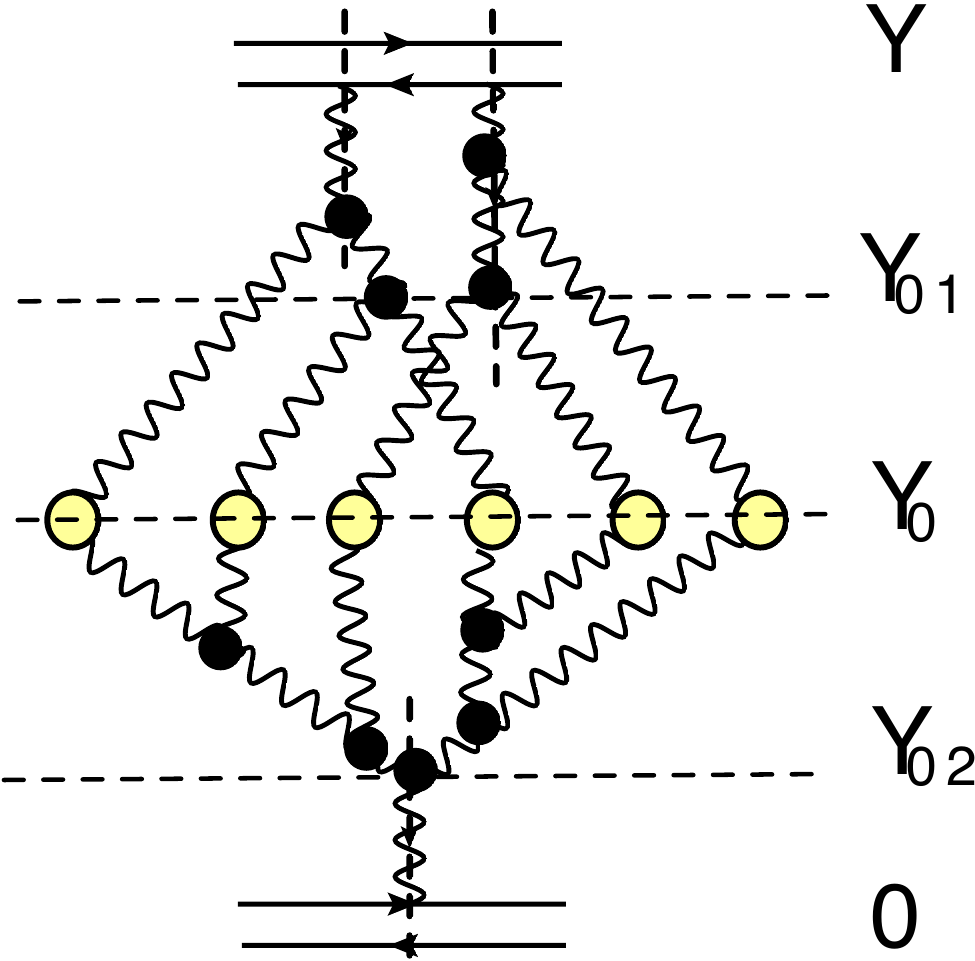} & ~~~~~~~~~& \includegraphics[width=5.7cm]{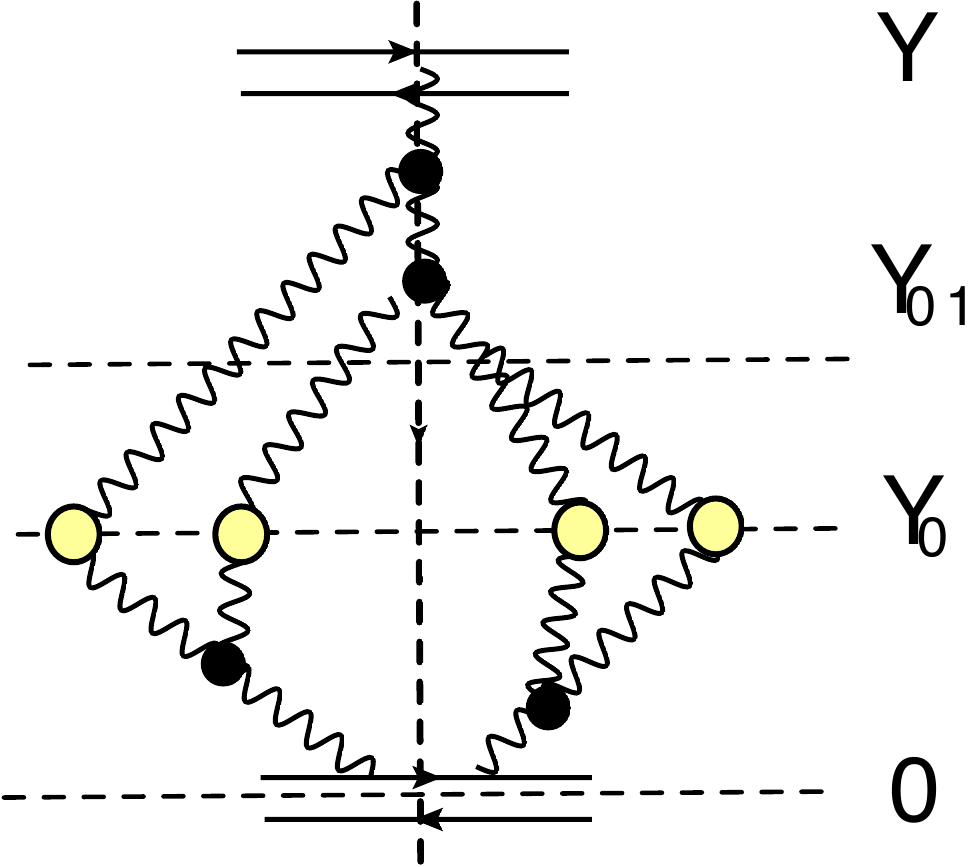}\\
     \fig{ddpom}-a &~~~~~~~~~& \fig{ddpom}-b\\
     \end{tabular}  
  \caption{ \fig{ddpom}-a: An example of  our calculation of double diffraction, using \eq{MPSI} for the Pomeron diagrams of \fig{sdpom}. \fig{ddpom}-b: An example of the diagrams for single diffraction in dipole-dipole scattering. The notations are the same as for \fig{sdpom}.}
\label{ddpom}
   \end{figure}

After finding $\rho_n$ for the fast dipoles we use \eq{MPSI} to find the double diffractive production cross section (see \fig{ddpom} in which we show an example of our calculations for  the simple diagram of \fig{sdpom}). However, as has been noted in Ref.~\cite{LEPRI}, we need to introduce two scattering amplitudes: $\gamma$ for the scattering amplitude in \fig{dodif} and $\bar{\gamma}$ for the complex conjugate amplitude (see \eq{PDD}). Bearing this in mind  \eq{DD17} takes the form:
  \begin{subequations} 
    \bea
\hspace{-1cm}S^D\Lb \Y,\Y_{01}\Rb &=&\frac{1}{1\,\,-\,\, e^{ \Y _m}\,\,+\,\, e^{ \Y_m}\, S^{-1}\Lb \gamma,\Y_{01}\Rb\,S^{-1}\Lb \bar{ \gamma},\Y_{01}\Rb}\,
=\,\frac{1}{1-\,\, e^{ \Y_m}\,+ e^{ \Y_m} \Lb 1+ \gamma_R\,e^{\Y_{01}}\Rb \Lb1 + \bar{\gamma}_R\,e^{\Y_{01}}\Rb}\label{DD18a}\\
&=& \frac{-1}{e^{ \Y_m}-1}\sum^\infty_{n=0} \Lb{\cal Z}\Lb\Y_m\Rb \Lb 1 + \gamma_R\,e^{\Y_{01}}\Rb\Lb1 + \bar{\gamma}_R\,e^{\Y_{01}}\Rb\Rb^n; \,\mbox{with}~~{\cal Z}\Lb \Y_m\Rb =\frac{e^{ \Y_m}}{e^{ \Y_m}-1};\label{DD18b}\\
&=& \frac{-1}{e^{ \Y_m}-1}\sum^\infty_{n=0} {\cal Z}^n\Lb \Y_m\Rb\sum^n_{l=0} \sum^n_{k=0}\frac{n!^2}{(n-l)!\,(n-k)!}\frac{ \Lb \gamma_R \,G_{\pom}\Lb \Y_{01}\Rb\Rb^l}{l!} \frac{  \Lb \bar{\gamma}_R\, G_{\pom}\Lb\Y_{01}\Rb\Rb^k}{k!}\label{DD18c} \eea
  \end{subequations} 
  Taking sum over $n$ in \eq{DD18c}  and taking into account  that $\rho_n\Lb\Y_{01}\Rb=C_nG_{\pom}^n\Lb\Y_{01}\Rb=\Lb\gamma_R G_{\pom}\Lb\Y_{01}\Rb\Rb^n$ (see \eq{DD16}) we rewrite \eq{DD18c} as
\beq \label{DD181}
  S^D\Lb \Y,\Y_{01}\Rb\,\,=\,\, \frac{-1}{e^{ \Y_m}-1} \sum^{\infty}_{l=0} \sum^{\infty}_{k=0}\frac{\Lb l!\Rb^2 {\cal Z}^l\Lb \Y_m\Rb \, _2F_1\Lb l+1,l+1;-k+l+1;{\cal Z}\Lb\Y_m\Rb\Rb}{(l-k)!}  \frac{\rho_l\Lb \Y_{01}\Rb}{l!} \frac{\rho_k\Lb \Y_{01}\Rb}{k!}
\eeq

  Recalling that $S^D$ is the diffractive S-matrix of the dipole  scattering with the nucleus at rest ($Y_0$ =0) 
 we can  use \eq{DD41} which has the following form for the one dimensional problem:
 \beq \label{DD1820}  
 \rho^A_n\Lb Y_0=0\Rb \,\,=\,\,\frac{\gamma_R^n}{n!}
\eeq  
Plugging \eq{DD1820} for the dipole densities in \eq{MPSID} with $Y_0 =0$ we can extract $\rho_{l,k}^D$ from $S^D$, viz.:
\beq \label{DD182}
  S^D\Lb \Y_m,\Y_{01}\Rb\,\,=\,\, \sum^\infty_{l=0}\,\sum^\infty_{k=0}\,\Lb -1\Rb^{l+k}\,\rho_{l,k}^D\Lb\Y_m,\Y_{01}\Rb\gamma^l_R\,\, \bar{\gamma}^k_R\eeq 
  Comparing \eq{DD181} with \eq{DD182} we obtain the following expression for the diffractive parton densities     $\rho^D_{l,k}\Lb \Y_m,\Y_{01}\Rb$:
  \beq \label{DD190}
  \rho^D_{l,k}\Lb \Y_m,\Y_{01}\Rb\,\,=\,\, \frac{\Lb -1\Rb^{-l-k+1}}{e^{ \Y_m}\,-\,1} \,\frac{l!\,{\cal Z}^l\,\Lb \Y_m\Rb \, _2F_1\Lb l+1,l+1;-k+l+1;{\cal Z}\Lb\Y_m\Rb\Rb}{(l-k)!\, k!}\,G^l_{\pom}\Lb\Y_{01}\Rb\,G^k_{\pom}\Lb\Y_{01}\Rb  \eeq
  where $ _2F_1\Lb l+1,l+1;-k+l+1;{\cal Z}\Rb$ is the hypergeometric function (see Ref.~\cite{RY}, formulae {\bf 9.2}).

  For large values of $\Y_m$,  ${\cal Z}\Lb\Y_m\Rb - 1 \,\to\,1/ G_{\pom}\Lb \Y_m\Rb \,\ll\,1$, and therefore $_2F_1\Lb l+1,l+1;-k+l+1;{\cal Z}\Lb\Y_m\Rb\Rb\,\simeq\,\frac{(l-k)!(l+k)!}{(l!)^2}\Lb 1- {\cal Z}\Lb\Y_m\Rb\Rb^{-l-k-1}$. In this limit, \eq{DD190} takes a simple form:
    \beq \label{DD20}
  \rho^D_{l,k}\Lb \Y_m,\Y_{01}\Rb\,\,=\,\, \frac{1}{e^{ \Y_m}-1} \frac{(l+k)!}{\Lb {\cal Z}( \Y_m)\,-\,1\Rb^{l+k+1}}   \frac{ G^l_{\pom}(\Y_{01})}{l!} \frac{  \, G^k_{\pom}(\Y_{01})}{k!}\,\Lb 1\,+\,{\cal  O}\Lb e^{ - \Y_m}\Rb\Rb.
  \eeq
     
   \eq{UT} equation takes a simple form in the one-dimensional model:
\beq \label{UT1}
G_{\pom} \Lb \Y\Rb \,\,=\,\,
G_{\pom}\Lb \Y - \Y_0\Rb G_{\pom}\Lb  \Y_0\Rb
\eeq
since $G_{\pom} \Lb Y\Rb $ is equal to $e^{ \Delta_{\pom} \,Y}$. Hence we can put $\gamma^{BA} = 1$ in \eq{MPSI}.   

Now we are ready to compute $S^{DD}$ in the one dimensional model, which takes the form
\beq \label{DD18BB}
S^{DD}\Lb \Y_{m1},\Y_{m2}; \Y_{01} - \Y_{02} \Rb\,=\,
    \,\sum^\infty_{l=0}\,\Lb -1\Rb^{l}\,l!\sum^\infty_{k=0}\,\Lb -1\Rb^{k}\,k!  \,\rho^{D}_{l,k}\Lb \Y_{m1} , \Y_{01} - \Y_0\Rb\,\rho^{D}_{l,k}\Lb \Y_{m2} ,\Y_0 - \Y_{02}\Rb
  \eeq 
  
Using this equation and \eq{DD20}
  one can see that $S^{DD}$ can be written as follows  from \eq{MPSIDD}  with $j=k + l$:
  
  \bea \label{DD19B}
 S^{DD}\Lb \Y_{m1},\Y_{m2}; \Y_{01} - \Y_{02} \Rb\,&= &\,\frac{1}{e^{ \Y_{m1}}-1}  \frac{1}{e^{ \Y_{m2}}-1} \sum^\infty_{j=0}\sum^j_{k=0}\frac{j!^2}{(j - k)!\,k!} \frac{\Lb - G_{\pom}\Lb \Y_{12}\Rb\Rb^j}{\Lb \Lb  {\cal Z}\Lb \Y_{m1}\Rb\,-\,1\Rb\, \Lb  {\cal Z}\Lb \Y_{m2}\Rb\,-\,1\Rb\Rb^{j+1}} \nn\\
 \,&= &\,\underbrace{\frac{1}{e^{ \Y_{m1}}-1}  \frac{1}{e^{ \Y_{m2}}-1} \frac{ 1}{\Lb  {\cal Z}\Lb \Y_{m1}\Rb\,-\,1\Rb\, \Lb  {\cal Z}\Lb \Y_{m2}\Rb\,-\,1\Rb}}_{=1}
\sum^\infty_{j=0} j!\,\Lb - \xi_{DD} \Rb^j\eea
where $ \xi_{DD} =\frac{ 2\, G_{\pom}\Lb \Y_{12}\Rb}{\Lb  Z( \Y_{m1})\,-\,1\Rb\, \Lb  Z( \Y_{m2})\,-\,1\Rb}$.
 
 Summing over $j$ we obtain:
  \beq \label{DD20B}
 S^{DD}\Lb \Y_{m1},\Y_{m2}; \Y_{01} - \Y_{02}   =  \Y_{12}\Rb\,\,=\,\,\frac{1}{\xi_{DD}} \,e^{\frac{1}{\xi_{DD}}}\,\Gamma\Lb 0, \frac{1}{\xi_{DD}}\Rb\,\xrightarrow{\xi_{DD} \,\gg\,1} \,\frac{\ln \xi_{DD}}{\xi_{DD}} 
 \eeq
 For large  $ \Y_{m1},\Y_{m2}$ and $\Y_{12} $ \eq{DD20B} leads to the following expression:
  \beq \label{DD21}
 S^{DD}\Lb \Y_{m1},\Y_{m2};\Y_{12}\Rb\,\,=\,\,\h \Lb  \Y_{m1} + \Y_{m2} +\Y_{12}\Rb\, e^{ - \Y_{m1} - \Y_{m2} - \Y_{12}}
 \eeq 

 For single diffraction in dipole-dipole scattering, \eq{MPSID} takes the following form:
 
  \beq \label{DD22}
      S^D_{dd}\Lb \Y_m,\Y_{01}\Rb\,=\,
    \,\sum^\infty_{n=0}\,\Lb -1\Rb^{n}\,n!\sum^\infty_{k=0}\,\Lb -1\Rb^{k}\,k!  \,\rho^{D}_{n,k}\Lb \Y_m , \Y_{01} - \Y_0\Rb\,\rho_n\Lb \Y_0\Rb \, \rho_k\Lb \Y_0\Rb   
      \eeq
 Plugging \eq{DD20} for $\rho^{D}_{n,k}\Lb \Y_m , \Y_{01} - \Y_0\Rb$ and \eq{DD16} for $\rho_n$ 
 and taking sum of $n+k$ we obtain
\beq \label{DD23}
      S^D_{dd}\Lb \Y_m, \Y_{01}\Rb\,\,=\,\,\frac{1}{\xi^2_D}\,e^{\frac{1}{\xi_D}}\,\Gamma\Lb -1,\frac{1}{\xi_D}\Rb \,\xrightarrow{\Y_m \gg 1} \,\frac{1}{\xi_D}\eeq
 where $ \xi_D =\frac{  G_{\pom}\Lb \Y_{01}\Rb}{\Lb  Z( \Y_{m})\,-\,1\Rb\,}$.  
 
\section{ Double diffraction in high energy QCD}
\label{DDIFhdQCD}
In QCD we need to find $\rho^D_{n,k}$ dipole densities to obtain the double diffraction cross section using \eq{MPSIDD}. As has been mentioned above, we start with diffractive production in dipole-nucleus scattering. The equation for single diffraction (see \eq{KL})
 for the scattering of a dipole with a nucleus  has the same form as the BK equation. The difference is only in the initial conditions of \eq{DD13} for $S^D$. Since BK equation in the saturation region shows the geometric scaling behaviour \cite{GS,IIML}, the solution to \eq{KL}  for the single diffraction can be generally written as follows~\cite{KOLE}:
\beq \label{DSOL1}
S^D\Lb z_m, z^{01}_d\Rb\,\,=\,\,S\Lb z_m\,\,+\,
\,\Delta z\Lb z^{01}_d\Rb\Rb;
\eeq
where function $\Delta z\Lb z^{01}_d\Rb$ is determined by the following equation:
\beq \label{DSOL2}
S\Lb \Delta z\Lb z^{01}_d\Rb\Rb\,\,=\,\, S^2\Lb z^{01}_d\Rb ~~\mbox{or} ~~~
\Omega\Lb \Delta z\Lb z^{01}_d\Rb\Rb\,\,=\,\, 2\, \Omega\Lb z^{01}_d\Rb
\eeq
with $S^D = \exp\Lb - \Omega^D\Rb$ and $S = \exp\Lb - \Omega \Rb$. The variables $z$, which we will use, are (see \fig{dodif}, \fig{mpsidd} and \fig{ddpom})
\beq \label{Z} 
 z\,=\, \,\kappa\,\bas Y  \,+\,\xi_{\vec{r}, \vec{R}};~~~~~~
 z^{01}_d\,=\, \,\kappa\,\bas\,Y_{01}  \,+\,\xi_{\vec{r}, \vec{R}};~~~~~~z^{12}_d\,\,=\,\,\kappa\,\bas\,\Lb Y_{01}\,-\,Y_{02} \Rb \,+\,\xi_{\vec{r}, \vec{r}'}
 \eeq
In \eq{Z}  the corresponding $\xi$'s  are determined in \eq{XI} of the Appendix~\ref{appendix:A}.


\subsection{Single diffraction for  the  leading-twist BFKL kernel} 


At the moment we are not ready to discuss the solution of   \eq{DSOL1} for the general BFKL kernel.   However, we are able to find a solution
 in the entire region of $z$ in the saturation region for 
  the simplified BFKL kernel suggested in Ref.~\cite{LETU}. This kernel describes the high-energy asymptotic solution of the nonlinear BK equation and leads to geometric scaling behaviour: 
    \bea \label{SIMKER}
\chi\Lb \gamma\Rb\,\,=\,\, \left\{\begin{array}{l}\,\,\,\frac{1} {1\,-\,\gamma}\,\,\,\,\,\,\,\,\,\,\mbox{for}\,\,z =  \kappa\,\bas\,Y+\xi_{\vec{r},\vec{R}} + \,>\,0,\,\,\,\,\,\,\mbox{summing} \Lb z\Rb^n;\\ \\
\,\,\,\frac{1}{\gamma}\,\,\,\,\,~~~~~~\mbox{for}\,\,\,z \,<\,0,\,\,\,\,\,~~~~~~\,\,\,\,\,~~~~~~~~~~~\mbox{summing}
\Lb \xi\Rb^n;\\  \end{array}
\right.
\eea
 where $\kappa = 4 $ for this kernel.
 
     \begin{figure}[ht]
    \centering
  \leavevmode
      \includegraphics[width=6cm]{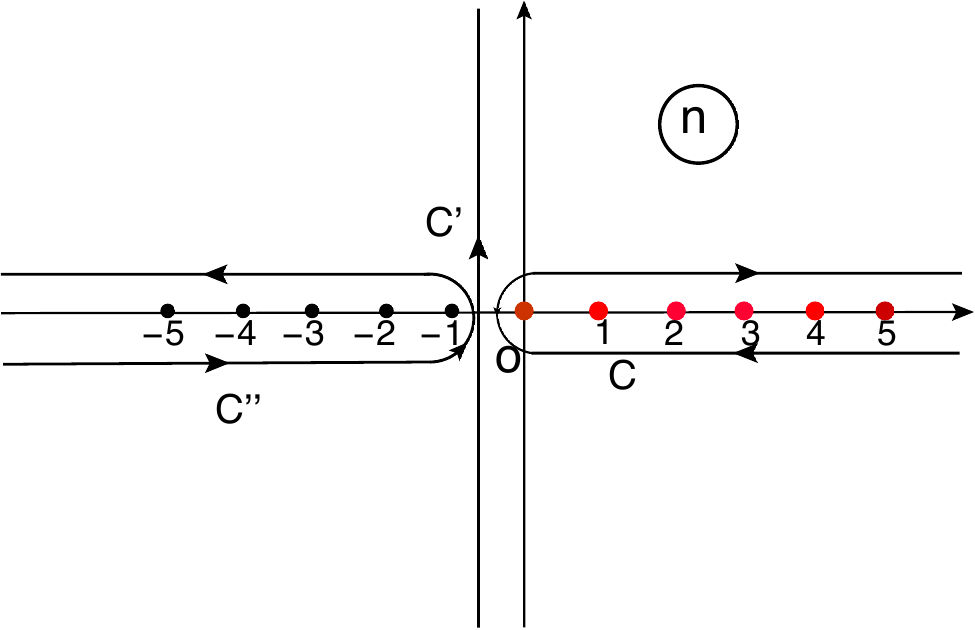}  
      \caption{The contours of integration in \eq{QCD5}. The red points denote the poles in $n$ for the right semi-plane. The black points show the poles in the left semi-plane where $S_d^d\Lb z\Rb$ has the double poles.}  
\label{cont}
   \end{figure}

 Since this kernel sums logarithmic contributions, it corresponds to the leading-twist term of the full BFKL kernel. It has a very simple form in the coordinate representation \cite{LETU}:
 
\beq \label{K2}
\intl \, \displaystyle{ K\Lb \vec{r}',\vec{r} - \vec{r'}|\vec{r}\Rb}\,d^2 r' \,\rightarrow
\,\frac{\bas}{2}\!\!\!\!\! \!\!\!\!\! \intl^{r^2}_{1/Q^2_s(Y,\vec{b})}\!\!\!\!\!\!  \frac{ d r'^2}{r'^2}\,\,+\,\,
\frac{\bas}{2}\!\!\!\!\! \!\!\!\!\! \intl^{r^2}_{1/Q^2_s(Y, \vec{b})}\!\!\!\!\!\!\! \frac{ d |\vec{r} - \vec{r}'|}{|\vec{r}  - \vec{r}'|^2}\,\,
 =\,\,\frac{\bas}{2}\, \intl^{\xi}_{\xi_s} d \xi_{\vec{r}'} \,\,+\,\,\frac{\bas}{2}\, \intl^{\xi}_{\xi_s} d \xi_{\vec{r} - \vec{r}'}\eeq
 where $\xi_{\vec{r}'} \,=\,\ln \Lb \frac{r'^2\,r^2_1}{b^4}\Rb$ for $b \,>\,r,r_1$
 and $\xi_s\,=\,\kappa\, \bas\,Y$ for the scattering of the  dipole $r'$ with the dipole $r_1$.
 Note that the logarithms
   originate from the decay of a large dipole into one small dipole and one large dipole~\cite{LETU}.  However, the size of the
 small dipole remains larger than $1/Q^2_s  = \frac{b^4}{r^2_1}\,e^{-\xi_s}  $.      Using  \eq{KL} we can obtain the nonlinear equation for  $\Omega^D\Lb \vec{x}_0,\vec{x}_1, Y,Y_{01}\Rb$, which determines the $S$-matrix for diffraction in dipole-nucleus interaction $S^D_{d A}\Lb  \vec{x}_0,\vec{x}_1, Y,Y_{01}\Rb=\exp\Lb - \Omega^D\Lb \vec{x}_0,\vec{x}_1, Y,Y_{01}\Rb\Rb$:
 \beq \label{BK1}
\frac{\partial^2\, \Omega^D\Lb \xi_s,\xi\Rb}{\partial \,\xi_s\,\,\partial\,\xi}\,\,=\,\,\frac{1}{4}\Lb 1\,\,-\,\,e^{ -\Omega^D\Lb \xi_s,\xi\Rb}\Rb
\eeq
The initial conditions for \eq{BK1}  are:
\beq \label{OMIC}
 \Omega^D\Lb  \vec{x}_0,\vec{x}_1, Y=Y_{01},Y_{01}\Rb = 2 \, \Omega\Lb  \vec{x}_0,\vec{x}_1, Y=Y_{01}\Rb
 \eeq
where $\Omega\Lb  \vec{x}_0,\vec{x}_1, Y=Y_{01}\Rb$ is defined through the elastic scattering matrix $S=e^{- \Omega}$.

\eq{BK1} has the geometric scaling solution~\cite{GS,IIML}  (see formula {\bf 3.5.3} in Ref.~\cite{MATH}), which has the following implicit form~\cite{CGML, CLMS}:
\beq \label{BK2}
\sqrt{2}\intl^{\Omega^D\Lb z_m,z^{01}_d \Rb}_{\Omega^D\Lb z_m=0, z^{01}_d\Rb} \frac{ d\,\Omega'}{\sqrt{ \Omega' -  1+ \exp\Lb - \Omega'\Rb}}\,\,=\,\,z_m\,\,\eeq
In \eq{BK2} $\Omega$ is the initial condition for $\Omega^D\Lb z_m=0 ,z^{01}_d \Rb$: $\Omega^D\Lb z_m =0,z^{01}_d\Rb = 2 \Omega\Lb z^{01}_d\Rb$, where the elastic $S$-matrix is equal to $S = e^{ -\Omega\Lb z^{01}_d\Rb}$.

Let us consider both $\Omega^D\Lb z_m,z^{01}_d \Rb$ and  $\Omega^D\Lb z_m =0,z^{01}_d\Rb$   being small. In this case \eq{BK2} reduces to
\beq \label{BK02A}
\sqrt{2}\intl^{\Omega^D\Lb z_m,z^{01}_d \Rb}_{\Omega^D\Lb z_m=0, z^{01}_d\Rb}\!\!\! \frac{ d\,\Omega'}{\sqrt{\h \Omega'^2 }}\,\,=\,\,z_m\,\,;~~~~~~~  \Omega^D\Lb z_m,z^{01}_d \Rb\,=\, \Omega^D\Lb z_m=0, z^{01}_d\Rb \,e^{\h \,z_m}\,;
  \eeq
 For small $z_m$ and $ z^{01}_d$ the main contribution to the scattering amplitude comes from the exchange of the BFKL Pomerons. Using \eq{DD11} and assuming the small values of $\Omega$'s we
 obtain: 
 \beq \label{BK02}  
 \Omega^D\Lb z_m=0, z^{01}_d\Rb \,\,=\,\,2\,N_0 \,e^{\h\,z^{01}_d} \,-\, N^2_0  \,e^{\,z^{01}_d}
 \eeq
 where $N_0$ is the value of the scattering amplitude for the BFKL Pomeron exchange, which we consider as being small.
 
 Finally, we have from \eq{BK02} that 
 \beq \label{BK002}
  \Omega^D\Lb z_m, z^{01}_d \Rb\,\,=\,\, \,\underbrace{2\,N_0\, e^{\h\,z}}_{\mbox{BFKL Pomeron exchange with $Y$}}\,-\,\underbrace{N_0^2 \,e^{\h z_m} \,e^{\,z^{01}_d} }_{ \mbox{Triple BFKL Pomeron contribution}}
 \eeq
 This confirms that \eq{BK2} correctly describes the Pomeron contributions to the scattering amplitude, and that all constants in the solution of \eq{BK1} have been chosen correctly.

Now we consider $\Omega\Lb z^{01}_d\Rb\,\gg\,1$. For such $\Omega\Lb z^{01}_d\Rb$, $\Omega'$ in \eq{BK2} turns out to be large and \eq{BK2} reduces to the following one:
\beq \label{BK20}
\sqrt{2}\intl^{\Omega^D\Lb z_m ,  z^{01}_d \Rb}_{2\,\Omega\Lb z^{01}_d\Rb} \frac{ d\,\Omega'}{\sqrt{ \Omega' }}\,\,=\,\,z_m\,;~~~~~~~~~~~~2\,\sqrt{2}  \Lb \sqrt{\Omega^D\Lb z_m,z^{01}_d \Rb}\,\,-\,\,\sqrt{2\,\Omega\Lb z^{01}_d\Rb}\Rb\,\,=\,\,z_m; \,\eeq
From this equation
\beq \label{BK21}
\Omega^D\Lb z_m,  z^{01}_d \Rb\,\,=\,\, \frac{1}{8} \,\Lb z_m \,+\,4\,\sqrt{\Omega\Lb z^{01}_d\Rb}\Rb^2 \,\,=\,\,\frac{1}{2\,\kappa} \,\Lb z_m \,+\,2\sqrt{\kappa\Omega\Lb z^{01}_d\Rb}\Rb^2 \,\,\eeq
Therefore, the solution of \eq{BK21} has the form of \eq{DSOL1} with $\Delta z\Lb z^{01}_d\Rb\,=\,\kappa  \sqrt{\Omega\Lb z^{01}_d\Rb}$. For large $ z^{01}_d$, $\Omega\Lb z^{01}_d\Rb= \frac{1}{2 \kappa}\Lb z^{01}_d\Rb^2 $ and \eq{BK21} takes the form:
\beq \label{BK22}
\Omega^D\Lb z_m, z^{01}_d \Rb\,\,=\,\,\frac{1}{8} \,\Lb z_m \,+\,\sqrt{2}  z^{01}_d\Rb^2
\eeq
with  $\Delta z\Lb z^{01}_d\Rb\,=\,\sqrt{2} z^{01}_d$. This solution shows that our \eq{DSOL1} is correct at least at large values of $z^{01}_d$.

\begin{figure}[ht]
    \centering
  \leavevmode
  \begin{tabular}{c c c}
     \includegraphics[width=5cm]{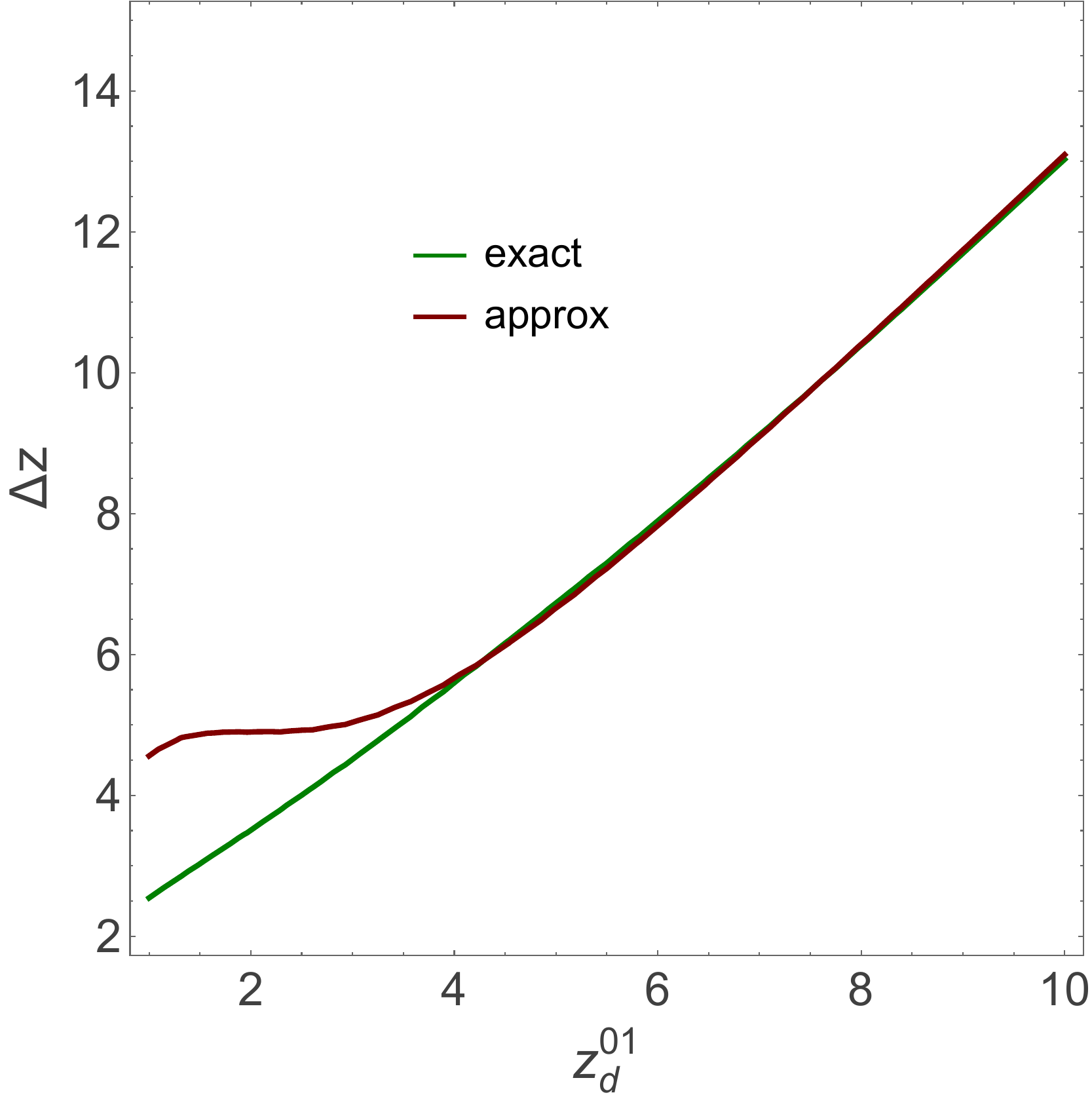} & ~~~~~~~~~& \includegraphics[width=5cm]{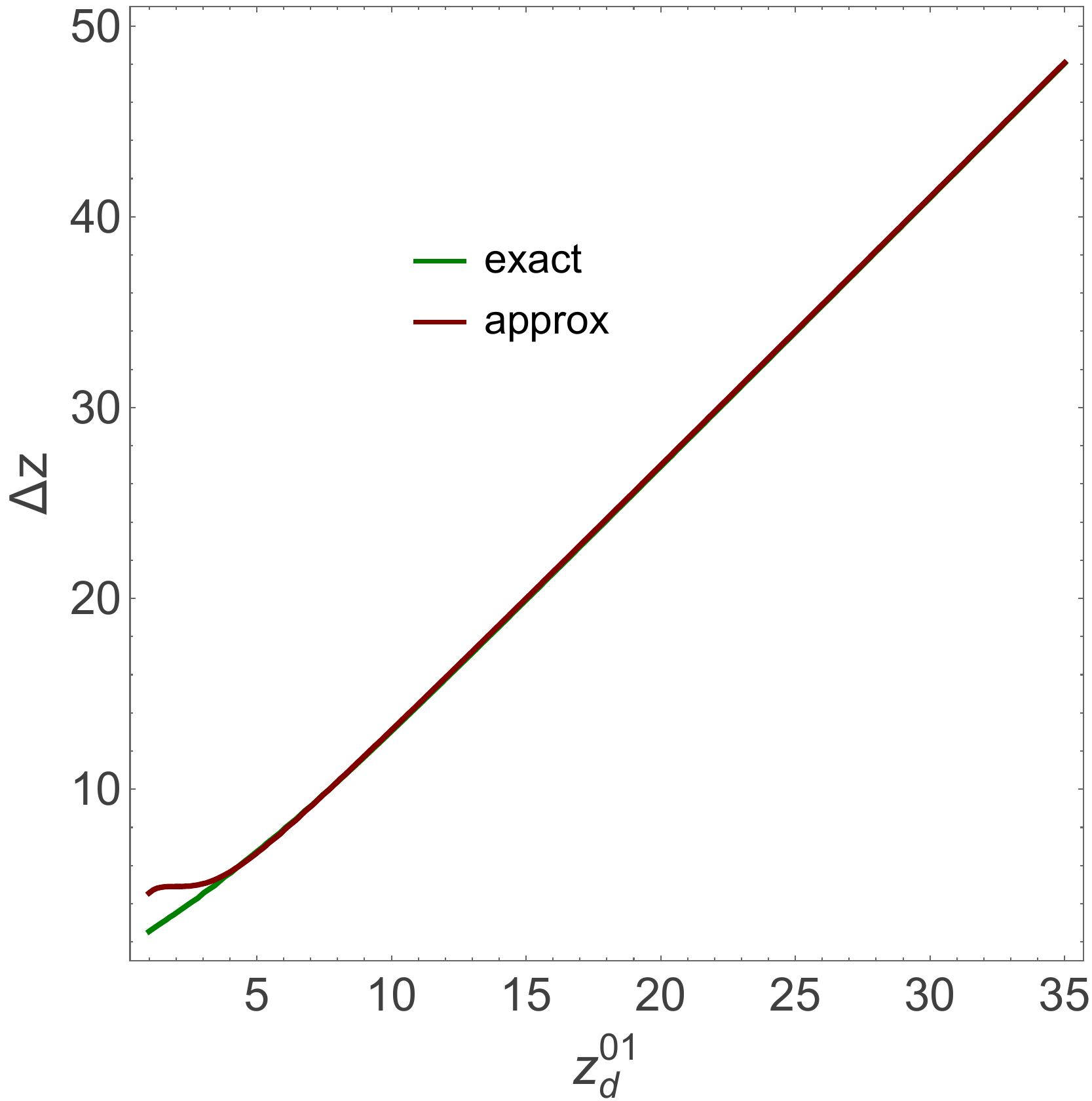}\\
     \fig{ddpom}-a &~~~~~~~~~& \fig{ddpom}-b\\
     \end{tabular}  
  \caption{ Function $\Delta z\Lb z^{01}_d\Rb$ versus $z^{01}_d$ for small  (\fig{dz}-a) and large (\fig{dz}-b) $z^{01}_d$.}
\label{dz}
   \end{figure}

Using solution of \eq{BK2} we can solve numerically \eq{DSOL2} and find function $\Delta z \Lb  z^{01}_d\Rb$. In \fig{dz} we show this solution: the curve labelled ``exact'' is the solution of \eq{DSOL2} with \eq{K2} for the elastic scattering amplitude, while $\Omega_{ approx }(z) = \frac{(z -4)^2}{8} +  \ln \Lb\h z\Rb$ is the approximate solution for the elastic amplitude taken from Ref.~\cite{LELT}. 

From \fig{dz} one can see that  the approximate $\Omega$ reproduces the function $\Delta z$ quite well for $z^{01}_d\,\geq\,4$, and for these values of $z^{01}_d$, $\Delta z\,\,\propto\,\,z^{01}_d$. 

Using \eq{DSOL1} and the series in \eq{QCD4} for the elastic $S$-matrix we rewrite $S^D$ in the form\footnote{Recall that for the leading-twist BFKL kernel  $\kappa$=4 and $\bg=\h$.}:

\beq \label{DSOL3}
S^D\Lb z_m, z^{01}_d\Rb = S\Lb z_m+\Delta z\Lb z^{01}_d\Rb\Rb =\sum_{n=0}^\infty (-1)^n C_n \,e^{ \h\,n\, \Lb z_m+\Delta z\Lb z^{01}_d\Rb\Rb}=\sum_{n=0}^\infty  \frac{(-1)^n}{n}\,e^{\frac{1}{2}\,n^2\,-\,2\,n} \,e^{ \h\,n\, \Lb z_m+\Delta z\Lb z^{01}_d\Rb\Rb}\eeq
with $C_n \,=\, \frac{1}{n}\exp\Lb \h \,n^2 \,-\, 2 \,n\Rb$ . In Appendix~\ref{appendix:B} we explain why $C_n \propto \exp\Lb \h n^2\Rb$. Other factors in \eq{DSOL3} have been discussed in Ref.~\cite{LELT}. They allow us to expand \eq{DSOL3} to the larger region of $z$'s.

It is worth mentioning that the leading-twist BFKL kernel allows us to prove that the nonlinear equation for diffraction has a solution that shows geometric scaling behaviour with the variable $  z_m \,+\,2\sqrt{\kappa\,\Omega\Lb z^{01}_d\Rb}$. It is instructive to recall that for the BK equation we also got the proof for the scaling behaviour of the scattering amplitude using the leading-twist BFKL kernel (see Ref.~\cite{LETU}).


\subsection{Double  diffraction for  the  leading-twist kernel} 


We need to introduce two kinds of dipoles in $S$ and $S^*$ (see  Sec.~\ref{DDIF1d} and \fig{mpsidd}) to find
$\rho^{D}_{n, k}(\vec{r}_1, \vec{b}_1\,
\ldots\,,\vec{r}_n, \vec{b}_n; \ldots\,,\vec{r}_k, \vec{b}_k; Y\,-\,Y_0,Y_{01})$ of \eq{MPSIDD}. Bearing this in mind we rewrite $S^D$ in the following form:
\beq \label{DSOL30}
S^D\Lb z_m, z^{01}_d\Rb=\sum_{n=0}^\infty  \frac{(-1)^n}{n}\,e^{\frac{1}{2}\,n^2\,-\,2\,n} 2^{-n} \sum_{k=0}^n \frac{n!}{k! \,(n -k)!} \,\,\underbrace{e^{ \h\,k\, \Lb z_m+\Delta z\Lb z^{01}_d\Rb\Rb}}_{\mbox{dipoles in  $S$} }\,\,\underbrace{e^{ \h\,(n-k)\, \Lb z_m+\Delta z\Lb z^{01}_d\Rb\Rb}}_{\mbox{dipoles in  $S^*$}}
\eeq

Using \eq{DSOL30}  and \eq{DD41} we can find  $\rho^{D}_{n, k}$ which take the form  (see notation in \fig{mpsidd}-b):
\bea \label{LTRHODD} 
&&\hspace{-0.5cm}\rho^{D}_{n, k}(Y\,-\,Y_0,Y_{01},\vec{r},\vec{b};\vec{r}_1, \vec{b}_1,\,
\ldots\,,\vec{r}_n, \vec{b}_n; \ldots\,,\vec{r}_k, \vec{b}_k)\,\,= \\
&&\,\,C_{n+k}\,2^{-n-k} \frac{(n+k)!}{n! \,k!} \underbrace{ \prod^n_{i=1}G_{\pom} \Lb z_{m1} +\Delta z\Lb  z^{01}_{i ,d}\Rb\Rb}_{\mbox{dipoles in  $S$} }\,\,\underbrace{\prod_{j=1}^k G_{\pom} \Lb z_{m1} +\Delta z\Lb z^{01}_{j ,d}\Rb\Rb}_{\mbox{dipoles in  $S^*$} }\,\nn  \eea

Plugging \eq{LTRHODD}  in \eq{MPSIDD} we obtain:
\bea \label{DDSOL1}
  &&  S^{DD}\Lb Y-Y_{01},Y_{02},Y_{01}-Y_{02}, \vec{r},\vec{r}' ;  \vec{b}\Rb\,=\\
 &&   \,\sum^\infty_{n=0}\,\Lb -1\Rb^{n}\,\sum^n_{k=0}\frac{n!^2}{k!(n-k)!}\   C^2_n \,2^{-2 n} G^{n}_\pom\Lb z_{m1} + \Delta z\Lb z^{01}_d \Rb  + z_{m2} +   \Delta z\Lb z^{02}_d \Rb\Rb \nn  \eea

As discussed above, $\Delta z \Lb z^{01}_d \Rb \,\propto\, z^{01}_d $; hence we can rewrite \eq{DDSOL1} in a more transparent form:
\bea\label{DDSOL2}
 S^{DD}\Lb Y-Y_{01},Y_{02},Y_{01}-Y_{02}, \vec{r},\vec{r}' ;  \vec{b}\Rb &=&
  \sum^\infty_{n=0}\,\Lb -1\Rb^{n}\,C^2_n 2^{-n} n! \,G^{n}_\pom\Lb z_{m1} + z_{m2} + \Delta z\Lb z^{12}_d \Rb  \Rb\\
  &=& \sum^\infty_{n=0}\Lb -1\Rb^{n}\,\frac{n!}{n^2}\,e^{ n^2 - (4+\ln 2) n} \,G^{n}_\pom\Lb z_{m1} + z_{m2} + \Delta z\Lb z^{12}_d \Rb \Rb  \nn   \eea

Rewriting the sum of $n$ as the counter integral as has been done in \eq{QCD9} and  \eq{QCD10}  and calculating the sum over poles using contour $C''$ (see \fig{cont}) we
obtain:
\beq \label{DDSOL3}
 S^{DD}\Lb z_{m1}, z_{m2} ,z^{12}_d\Rb\,\,=\,\,
\oint_{C''}\frac{d \,n}{2\,\pi\,i}   \,   \frac{\pi\,\Gamma\Lb 1+n\Rb}{\sin\Lb \pi \,n\Rb} \frac{1}{n^2}e^{ n^2 -(4+\ln2)\, n } e^{\h\,n\,\Lb z_{m1} +z_{m2}+ \Delta z\Lb z^{12}_d \Rb  \Rb}  \eeq
    
 Using $\Gamma(1-x)\,\Gamma(x)=\frac{\pi}{\sin(\pi x)}$ we  reduce \eq{DDSOL3} to the form:
 \bea \label{DDSOL4}
 &&S^{DD}\Lb z_{m1}, z_{m2} ,z^{12}_d\Rb\,\,=
\oint_{C''}\frac{d \,n}{2\,\pi\,i}\frac{\pi^2}{\sin^2\Lb \pi \,n\Rb\,\Gamma(-n)}\,\frac{1}{n^2}\,e^{ n^2 - (4+\ln2) n } e^{\h\,n\,\Lb z_{m1} +z_{m2}+ \Delta z\Lb z^{12}_d \Rb  \Rb}\,\,\\
&&\hspace{-0.8cm}  =\,\,\sum_{k=1}^{\infty} \,\frac{1}{k^2\,\Gamma(k)}\,\,\underbrace{\Lb \psi ^{(0)}(k)\,-\, 2\,k \,+\,\frac{2}{k}\,-\,(4+\ln 2)\,+\, \h\, \Lb z_{m1} +z_{m2}+ \Delta z\Lb z^{12}_d \Rb\Rb\Rb}_{double\,\, pole\,\,residue}\, \,e^{ k^2 \,+\, (4+\ln 2) k \,-\,\h\,k\,\Lb z_{m1} +z_{m2}+ \Delta z\Lb z^{12}_d \Rb\Rb} \nn
\eea   
    
    Repeating all the steps of the calculation (see \eq{QCD9})  with the saddle point at large $k$:  $k_{SP}=\frac{1}{4}\Lb z_{m1} +z_{m2}+ \Delta z\Lb z^{12}_d\Rb\Rb\,-\,\h(4+ \ln2)$,  we obtain the following answer:
    
\beq \label{DDSOL5}
 S^{DD}\Lb z_{m1}+z_{m2} +\Delta z\Lb z^{12}_d\Rb\Rb\,=\,\,\frac{\Phi\Lb  z_{m1}+z_{m2} +\Delta z\Lb z^{12}_d\Rb \Rb}{\,\Gamma\Lb \frac{ z_{m1}+z_{m2} +\Delta z\Lb z^{12}_d\Rb }{4}\,-\,\h(4 + \ln 2)\Rb}\,\exp\Lb -\, \frac{ \Lb  z_{m1}+z_{m2} +\Delta z\Lb z^{12}_d\Rb\,\,-\,\,8\,-\,2\,\ln 2\Rb^2}{16}\Rb\eeq
where $\Phi$ is a smooth function of $  n_{SP}$. In \eq{DDSOL5} this function is equal to $\Phi\Lb k_{SP}\Rb-\frac{\sqrt{\pi}}{k^2_{SP}}\Lb \psi^{(0)}\Lb k_{SP}\Rb + \frac{2}{k_{SP}}\Rb$. 

It is easy to see that $S^{DD}$ of \eq{DDSOL5} satisfies the following initial and boundary conditions:
 \begin{subequations}   
\bea 
S^{DD}\Lb z_{m1} , z_{m2}=0, z^{12}_d\Rb\,&=&\,S^D_{dd}\Lb z_{m1},z^{12}_d\Rb 
;\label{QCDDD2}\\
S^{DD}\Lb z_{m1}=0 , z_{m2}, z^{12}_d\Rb\,&=&\,S^D_{dd}\Lb z_{m2},z^{12}_d\Rb ;\label{QCDDD3}\\
S^{DD}\Lb z_{m1}=0, z_{m2}=0 , z^{12}_d\Rb\,&=&\,\Lb S^d_d\Lb z^{12}_d \Rb\Rb^2;\label{QCDDD4}
\eea
\end{subequations}


\subsection{Double  diffraction for  the full BFKL kernel} 

In the previous  section we worked with the simplified leading-twist BFKL kernel. It is instructive to note, however, that \eq{DSOL1} and \eq{DSOL2} can be generalized to double diffraction for the regular BFKL kernel. Indeed, 

\beq \label{DDSOL6} 
S^{DD}\Lb z_{m1}+z_{m2} +\Delta z\Lb z^{12}_d\Rb\Rb\,\,=\,\,S^d_d\Lb z_{m1}+z_{m2} +\Delta z\Lb z^{12}_d\Rb\Rb
\eeq
where $S^d_d$ is the scattering matrix for dipole-dipole interaction. This solution has the initial conditions of \eq{QCDDD2}-\eq{QCDDD4} as well as \eq{DDSOL5} for the leading-twist BFKL kernel. $\Delta z \Lb z^{12}_d\Rb$ is the solution of the following equation:
\beq \label{DDSOL7}
S\Lb \Delta z\Lb z^{12}_d\Rb\Rb\,\,=\,\, S^2\Lb z^{12}_d\Rb ~~\mbox{or} ~~~
\Omega\Lb \Delta z\Lb z^{12}_d\Rb\Rb\,\,=\,\, 2\, \Omega\Lb z^{12}_d\Rb
\eeq

\begin{figure}[ht]
    \centering
  \leavevmode
      \includegraphics[width=9cm]{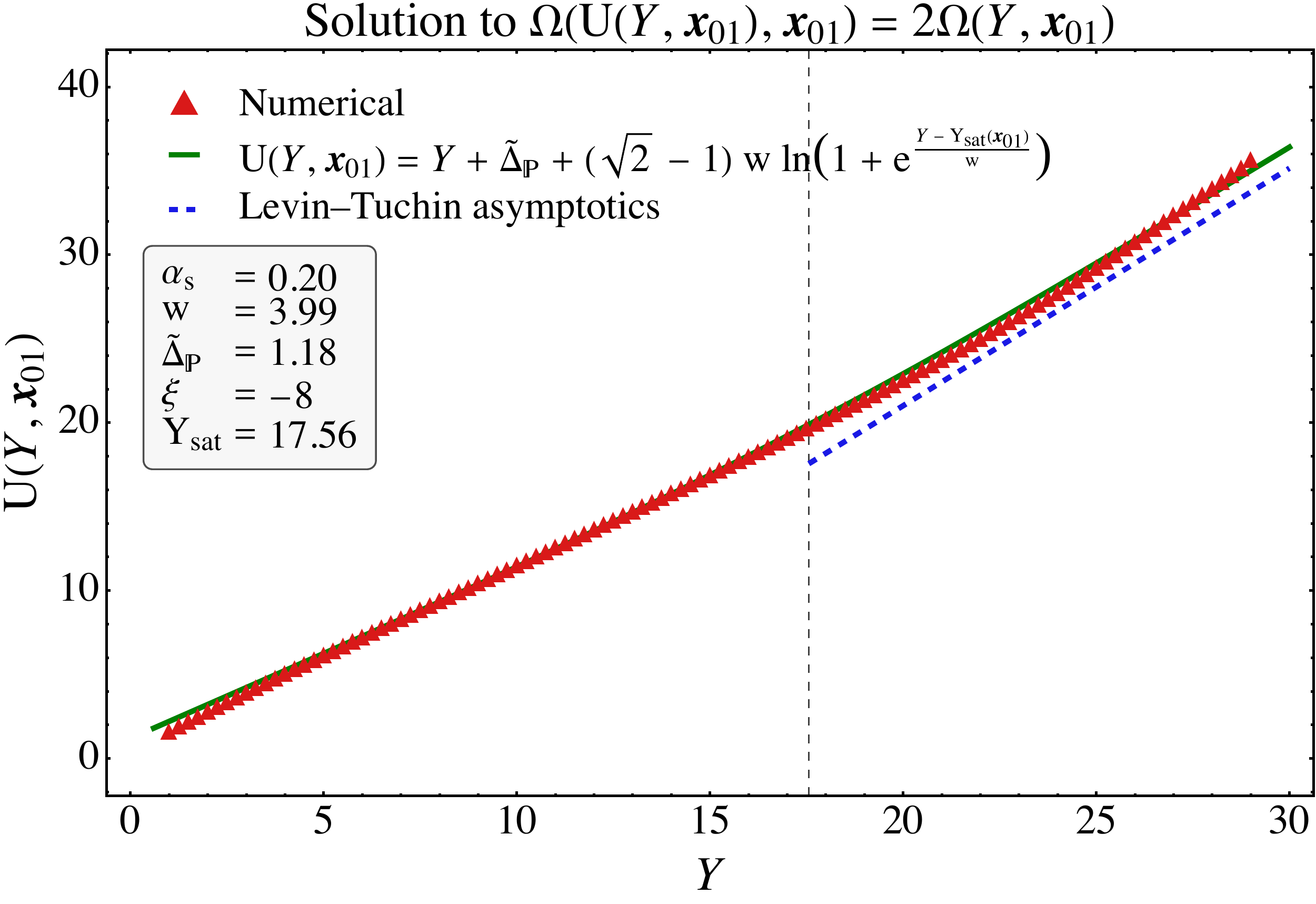}  
      \caption{Function $U\Lb Y, \vec{x}_{01}\Rb$ at fixed $\xi  =  \ln |\vec{x_{01}}|^2$ = -8 (see \eq{DDSOL71}). The parametrization is given by \eq{DDSOL72}. The initial conditions are given by McLerran--Venugopalan formula~\cite{MV} with $Q_s(Y=Y_A) =1$ GeV.}   
\label{u}
   \end{figure}

This equation is  the ${\rm  Schr\ddot{o}der}$-type doubling equation.
Unfortunately we cannot solve \eq{DDSOL7} analytically  in the entire kinematic range of the variable, but we found a numerical solution for the equation:
\beq \label{DDSOL71}
\Omega\Lb U\Lb Y, \vec{x}_{01}\Rb, \vec{x}_{01} \Rb\,\,=\,\, 2\, \Omega\Lb Y,\vec{x}_{01}\Rb
\eeq
which is plotted in \fig{u}. The numerical solution for $U\Lb\Lb Y, \vec{x}_{01}\Rb, \vec{x}_{01} \Rb$ can be parameterize as
\beq \label{DDSOL72}
 U\Lb Y, \vec{x}_{01}\Rb\,\,=\,\,Y\,+\tilde\Delta_{\pom} + (\sqrt{2}\,-\,1)\,w\,\ln \Lb 1\,+\,\exp\Lb \Lb Y - Y_{sat}\Lb \vec{x}_{01}\Rb\Rb/w\Rb\Rb 
\eeq
where $\ln Q^2_s\Lb Y_{sat}\Rb = \ln\Lb 1/x^2_{01}\Rb$, $\tilde\Delta_{\pom}=(\ln 2)/\Delta_{\pom}$ and $w=3.99$ is the only parameter which was fitted to the data. From \fig{u} one can see that \eq{DDSOL72} describe the solution with accuracy about 11\%. The main lesson from this equation is that  the solution reproduces the large Y limit for $Y \geq Y_{sat}$.

\begin{figure}[ht]
    \centering
  \leavevmode
  \begin{tabular}{c c}
      \includegraphics[width=8.5cm]{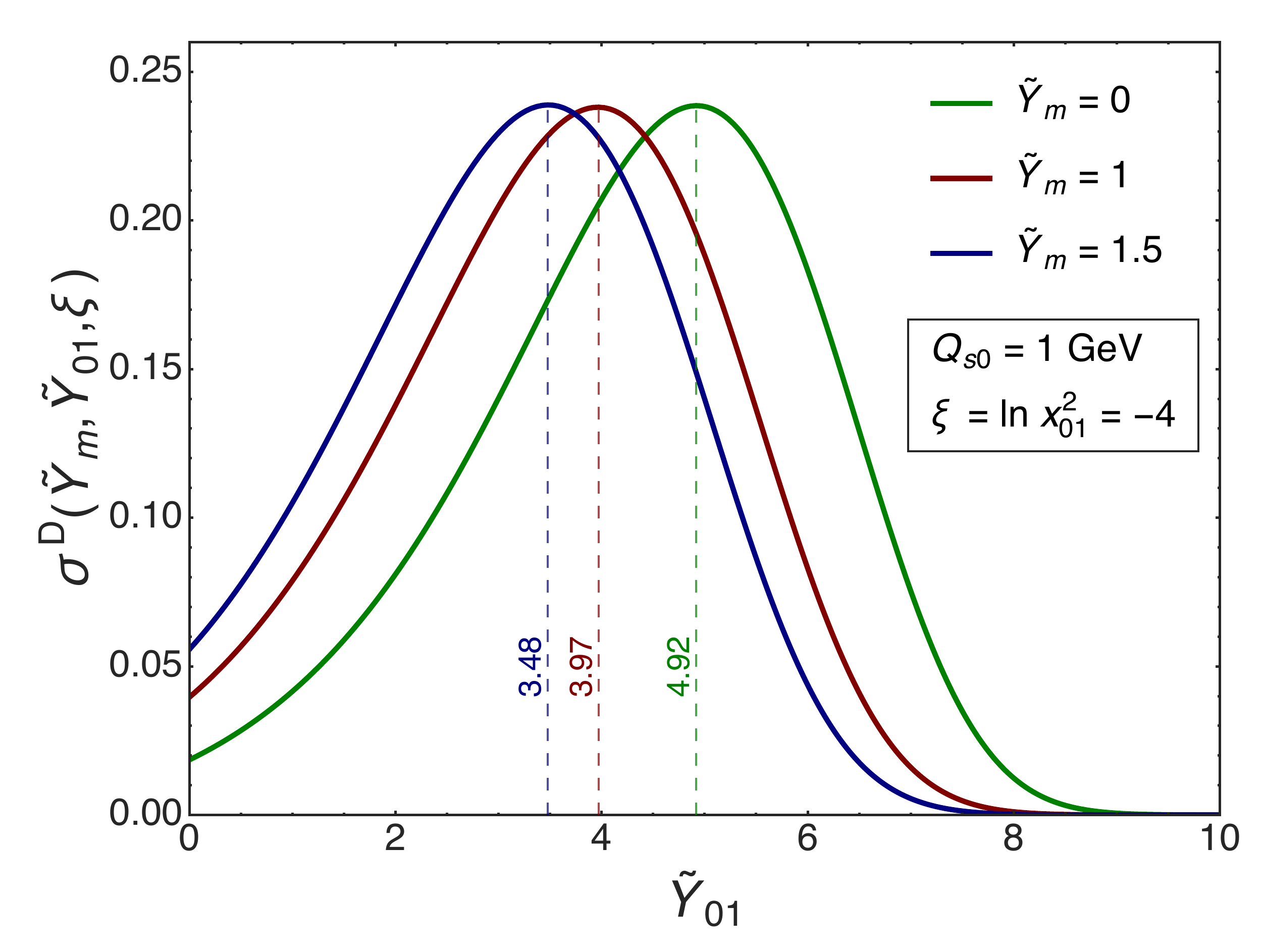} &  \includegraphics[width=8.5cm]{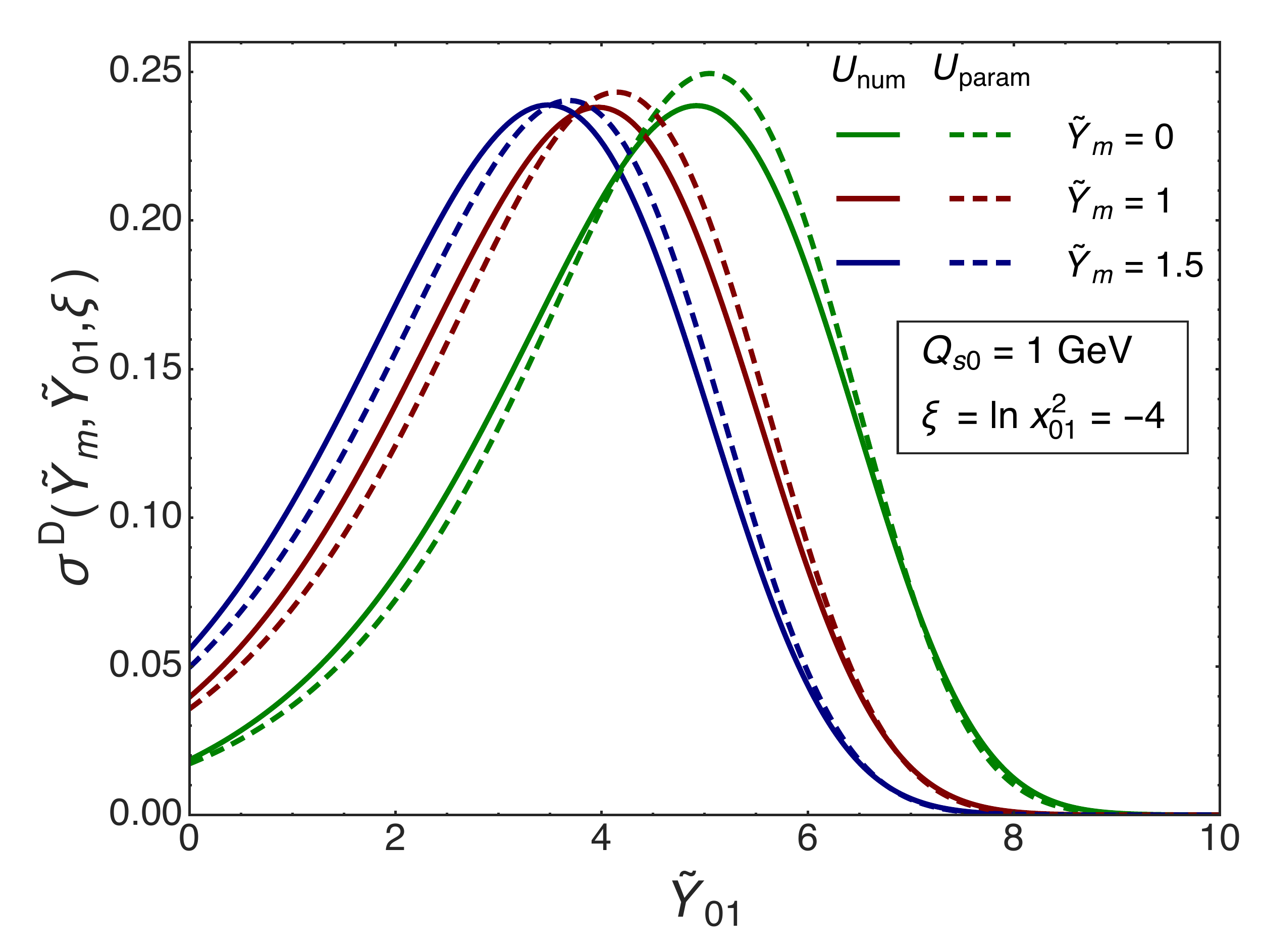}      \\
  \fig{sdfker}-a &    \fig{sdfker}-b\\
     \end{tabular} 
      \caption{$\sigma^D(\Y_m,\Y_{01}) = -\frac{\pp S^D(\Y_m,\Y_{01})}{\pp \Y_{01}}$ versus $\Y_{01}$ at fixed $\Y_m$ and fixed $\xi=\ln|\vec{x}_{01}|^2 = - 4$. \fig{sdfker}-a: Numerical solution to the KL equation. \fig{sdfker}-b: Numerical solution to the KL equation with $U\Lb \Y, \xi\Rb$ for parameterization of \eq{DDSOL72}. The initial conditions are given by McLerran--Venugopalan formula~\cite{MV} with $Q_s(Y=Y_A) =1$ GeV. The rescaled rapidities $\Y_m=\Delta_{\pom}Y_m$ ($\Y_{01}=\Delta_{\pom}Y_{01}$) are analogous to those defined in Section~\ref{DDIF1d} for the one-dimensional model.}\label{sdfker}
   \end{figure}

 It is instructive to notice, that $\sigma^D = - \pp S^D/\pp Y_{01}$ (see \eq{SD1}) shows the same main features as in the one dimensional model(see \fig{xsdif} ):
 each curve is a smooth bump. The peak sits at $Y_{01} \,\approx  9.3,\ 8.3,\ 7.9$ for $Y_m=0,\ 1,\ 1.5$\footnote{Note that in \fig{sdfker} we used a different  Y variable: $\Y= \Delta_{\pom}\, Y$} at $\alpha_S=0.2$, so a larger $Y_m$ moves the peak to smaller $Y_{01}$ (see \fig{sdfker}-a). The value of $\sigma^D$ at the peaks are the same in contrast with the one dimensional model (see \fig{xsdif}). It worthwhile  mentioning that for the parameterization of \eq{DDSOL72} the values of $\sigma^D$ in the peaks changes.

 However, we are able to find the analytic  solution in the region where $z_{m1}$, $z_{m2}$ and $z^{12}_d$ are large.
 In this region this equation reduces \cite{LETU} to the  simple linear evolution equation:
\beq \label{QCD1}
  \frac{\partial  S^{D}\Lb \vec{x}_1,\vec{x}_0, Y_m , Y_{01}\Rb}{\partial Y_m}\,\,=\,\,-
  \,\frac{\bas}{2\,\pi}\int d^2\,x_{02}\, \frac{ x^2_{01}}{x^2_{02}\,x^2_{12}}\,S^{D}\Lb \vec{x}_1,\vec{x}_0, Y_m , Y_{01}\Rb 
  \eeq
In the saturation region where we have the geometric scaling behaviour~\cite{GS,IIML}, \eq{QCD1} takes the simple form:
\beq \label{QCD2}
 \kappa \frac{d\, S^{D}\Lb z_m, z^{01}_d\Rb}{d\,z_m}\,\,=\,\,-z_m\,\,
S^{D}\Lb z_m, z^{01}_d\Rb\eeq
where $r$ and $R$ are the sizes of incoming dipoles and the dipoles inside of the nucleus. The variables  $z_m$ and $  z^{01}_d $ have been defined in \eq{Z}. For $\Omega\Lb z_m, z^{01}_d\Rb$ \eq{QCD2} takes the form:
\beq \label{QCD20}
 \kappa \frac{d^2\, \Omega^{D}\Lb z_m, z^{01}_d\Rb}{d\,z_m^2}\,\,=\,\,1\,\,
\eeq
The general solution of this equation is
\beq \label{QCD21}
 \Omega^{D}\Lb z_m, z^{01}_d\Rb \,\,=\,\,\frac{1}{\kappa}\Lb \h z^2_m \,+\, f(z^{01}_d)\,z_m \,+\,  p(z^{01}_d) \Rb
 \eeq
 Functions $f$ and $p$ have to be found from the initial conditions:  $\Omega^D\!\!\Lb z_m =0,z^{01}_d\Rb\,=\,2\,\Omega\Lb z^{01}_d\Rb $, where $\Omega$ describes the elastic scattering. One can see that
 \beq \label{QCD22}
 f(z^{01}_d)\,\,=\,\,2 \sqrt{\,\kappa\,\Omega\Lb z^{01}_d\Rb};~~~~~ p(z^{01}_d) \,\,=\,\,2\,\kappa\,\Omega\Lb z^{01}_d\Rb ;\eeq 
 leads to
 \beq \label{QCD23}
  \Omega^{D}\Lb z_m, z^{01}_d\Rb\,\,=\,\, 
 \Omega^{D}\Lb z_m \,+\,2\,\sqrt{\kappa\,\Omega\Lb z^{01}_d\Rb} \Rb =\frac{1}{2\,\kappa}\Lb  z_m  + 2\,\sqrt{\kappa\,\Omega\Lb z^{01}_d\Rb} \Rb^2 \eeq 
 This solution satisfies the initial condition for $\Omega^D$.
 
For large $z^{01}_d$, $\Omega\Lb z^{01}_d\Rb\,=\,  \frac{1}{2\,\kappa}\Lb z^{01}_d \Rb^2$ reducing   \eq{QCD23} to  the form:
\beq \label{QCD3}
S^D\Lb z_m,z^{01}_d\Rb\,\,=\,H\Lb \frac{\Lb z_{m}   \,+\, \sqrt{2} z^{01}_d\Rb^2}{2\,\kappa}\Rb \,\exp\Lb  - \frac{\Lb z_{m}   \,+\, \sqrt{2} z^{01}_d\Rb^2}{2\,\kappa}\Rb ~~~\mbox{with}~~~z_m\,=\, \,\kappa\,\bas\,Y_{m} \,+\,\xi_{r, R}
\eeq
where $H$ is a smooth function which leads to a small correction to \eq{QCD2}. 
In \eq{QCD3} we use the initial condition of \eq{DD13} and assume that $z_d^{01}$ is large and we have the geometric scaling behaviour for the elastic $S$.
We have thus shown that at large values of $z$ we obtain the same solution as for the leading-twist BFKL kernel. The double diffraction therefore takes the form:
\beq \label{DDSOL7B}
 S^{DD}\Lb z_{m1}+z_{m2} +\Delta z\Lb z^{12}_d\Rb\Rb\,=\,H\Lb  z_{m1}+z_{m2} +\Delta z\Lb z^{12}_d\Rb \Rb\,\exp\Lb -\, \frac{ \Lb  z_{m1}+z_{m2} +\Delta z\Lb z^{12}_d\Rb\,\,-\,\,8\,-\,2 \ln 2\Rb^2}{4\,\kappa}\Rb
\eeq
with $\Delta z \Lb z^{12}_d\Rb\,=\,\sqrt{2} z^{12}_d$. This solution is shown in \fig{u}   by dotted line.

\subsection{Possible relation to the ``rare" fluctuations approach} 
In this section, we present our results in a form close to that of the "rare" fluctuation approach~\cite{IAMU}. This allows us to discuss how the two approaches relate to each other.

\subsubsection{Elastic dipole-dipole scattering} 


 Let us consider the scattering of a dipole $r$ at rapidity $Y$ with the dipole $r_d$ at rapidity $Y_0$ in the BK approach. This amplitude satisfies the BK equation  and can be written as the following series \cite{LELU}\footnote{All notations are  the same as in \eq{MPSI}-\eq{QS}.}:
\beq \label{DDEQ1}
S\Lb Y-Y_0, \vec{r}, \vec{b}, \vec{r}_d,\vec{b}_d \Rb\,=\,\sum_{n=0}^\infty (-1)^n \intl \prod^n_{i=1 } d^2 r_i\,d^2 b_i\,\gamma^{BA}\Lb \vec{r}_i,\vec{b}_i, \vec{r}_d,\vec{b}_d\Rb \rho_{n}\Lb Y-Y_0,\vec{r},\vec{b};\{\vec{r}_i,\vec{b}_i\}\Rb \,=\,\sum^\infty_{n=0} \Lb - 1 \Rb^n C_n\,e^{\,n\,\bg\, z_d}
\eeq
where $z_d \,\,=\,\,\bas\,\kappa \Lb Y - Y_0\Rb + \xi_{\vec{r},\vec{r}_d}$. In the second equation we use the features of the BFKL Pomeron in the saturation region that has been discussed in \eq{UT}. For the geometric scaling solution the coefficients $C_n$ can be found as follows:
\beq \label{DDEQ2}
(-1)^n\,C_n\,\,=\,\,\intl^{\epsilon+i \infty}_{\epsilon-i \infty}\,\frac{d \,\bg z_d}{2\,\pi\,i}\,e^{ - n\,\bg\,z_d}\,S\Lb z_d\Rb\,\eeq

For the scattering amplitude of  the dipole $r_d$ at rapidity $Y_0$ with the dipole $r'$ at rest we have \eq{DDEQ2} in the form:
\beq \label{DDEQ3}
(-1)^n\,C_n\,\,=\,\,\intl^{\epsilon+i \infty}_{\epsilon-i \infty}\,\frac{d \,\bg \,\bar{z}_d}{2\,\pi\,i}\,e^{ - n\,\bg\,\bar{z}_d}\,S\Lb \bar{z}_d\Rb\eeq
with $\bar{z}_d \,\,=\,\,\bas\,\kappa\,Y_0 \,+\, \xi_{\vec{r}_d,\vec{r}'}$.

Using these equations we can rewrite \eq{QCD8} for $S^d_d$ as follows:

\beq \label{DDEQ4}
S^d_d\Lb z^{12}_d\Rb\,\,=
\,\,\sum_{n=0}^{\infty} (-1)^n\,C^2_n\,\Gamma\Lb n+1\Rb e^{n\,\bar{\gamma}\, z^{12}_d}=\sum_{n=0}^{\infty} (-1)^n\,n!\,\,\intl^{\epsilon +i \infty}_{\epsilon-i \infty} \frac{d \,\bg\,z_d}{2\,\pi\,i}\intl^{\epsilon +i \infty}_{\epsilon-i \infty} \frac{d \,\bg\,\bar{z}_d}{2\,\pi\,i}e^{ n\,\bg\, \Lb  z^{12}_d - z_d - \bar{z}_d\Rb} S\Lb z_d\Rb\,S\Lb\bar{z}_d\Rb\eeq

First, let us neglect the factor $(-1)^n\,n!$ in \eq{DDEQ4}. This simplification gives:
\beq \label{DDEQ5}
S^d_d\Lb z^{12}_d\Rb\,\,=\,\,\sum_{n=0}^{\infty} \,\,\intl^{\epsilon +i \infty}_{\epsilon-i \infty} \frac{d \,\bg\,z_d}{2\,\pi\,i}\intl^{\epsilon +i \infty}_{\epsilon-i \infty} \frac{d \,\bg\,\bar{z}_d}{2\,\pi\,i} e^{ n\,\bg\, \Lb  z^{12}_d - z_d - \bar{z}_d\Rb}\,S\Lb z_d\Rb\,S\Lb\bar{z}_d\Rb
\,\,\to\,\,\intl^{\epsilon +i \infty}_{\epsilon-i \infty} \frac{d \,\bar{z}_d}{2\,\pi\,i} S\Lb z^{12}_d - \bar{z}_d\Rb\,S\Lb\bar{z}_d\Rb
\eeq
We can easily take into account  $(-1)^n\,n!$   
considering  \eq{DD21} and the definition of the Borel transform~\cite{BOREL} (see also Appendix~\ref{appendix:C}):
\beq \label{DDEQ6}
S\Lb e^{\bg\,z}\Rb\,\,=\,\,\intl^\infty_0\!d t \,e^{-t} {\cal B}S\Lb - t e^{\bg\,z}\Rb
\eeq
Using this equation and \eq{DDEQ5} we obtain for the scattering matrix  our master equation:
\beq \label{DDEQ7}
S^d_d\Lb z^{12}_d\Rb\,\,=\,\,\intl^\infty_0 d t \,e^{-t}\, \frac{1}{\bg}\intl^{\epsilon +i \infty}_{\epsilon-i \infty} \frac{d \,\bar{z}_d}{2\,\pi\,i}\underbrace{ {\cal B}S\Lb \ln(- t)/\bg + z^{12}_d - \bar{z}_d\Rb}_{\mbox{BK equation}}\,\,\underbrace{S\Lb\bar{z}_d\Rb}_{\mbox{BK equation}}\eeq
First, we check how \eq{QCD10} appears from this equation. In the saturation regions for both ${\cal B}S\Lb \ln (-t)/\bg + z^{12}_d - \bar{z}_d\Rb$ and $S\Lb\bar{z}_d\Rb$ we reduce the master equation to the following integral:
\beq \label{DDEQ8}
S^d_d\Lb z^{12}_d\Rb\,\,=\,\,\intl^\infty_0 d t \,e^{-t}\, \frac{1}{\bg}\intl^\infty_{- \infty} \frac{d \,\bar{z}_d}{2\,\pi} \,\exp\Lb - \frac{\Lb  \ln(- t)/\bg + z^{12}_d - \bar{z}_d\Rb^2 + \bar{z}^2_d}{2 \,\kappa}\Rb 
\eeq

Taking the integral over $\bar{z}_d $ using the method of steepest descent with the saddle point $\bar{z}_{d,SP}  = \h\Lb \ln (-t)/\bar\gamma + z^{12}_d \Rb$ we reproduce \eq{QCD10}:
\beq \label{DDEQ9}
S^d_d\Lb z^{12}_d\Rb\,\,=\,\,\frac{\sqrt{\kappa}}{2 \sqrt{\pi}\,\bg}\intl^\infty_0 d t\, e^{-t} \,\exp\Lb - \frac{\Lb \ln(- t)/\bg + z^{12}_d\Rb^2}{4 \,\kappa} \Rb
\xrightarrow{z^{12}_d \gg 1} H\Lb \frac{z^{12}_d}{2\,\kappa\,\bg}\Rb \Gamma\Lb 1 - \frac{z^{12}_d}{2\,\kappa\,\bg}\Rb\,\exp\Lb - \frac{\Lb z^{12}_d\Rb^2}{4\,\kappa}\Rb
\eeq
where $H$ is a smooth function of its argument. More detailed calculations can be found in Appendix~\ref{appendix:C}.

These estimates can be interpreted in a way that is close to the ``rare" fluctuation estimates of Ref.~\cite{IAMU}. Indeed, \eq{DDEQ7} has the physical meaning of a ``rare" fluctuation of size $Y$: it is the product of two probabilities, the probability of a fluctuation of size $Y-Y_0$ in the BK equation for the upper amplitude, and the probability of a small fluctuation of size $Y_0$ for the lower amplitude. The equation itself is the sum over all such possible small fluctuations, and \eq{DDEQ9} reproduces the same result as Ref.~\cite{IAMU}. In this discussion we have used the interpretation of the BK solution at large rapidities \cite{LETU} as a ``rare"  fluctuation, following Ref.~\cite{IAMU}. We believe that \eq{DDEQ7}, together with its connection to the ``rare"  fluctuation approach, will allow us to estimate the range of validity of both approaches, since summing the large Pomeron loops can only be trusted for the restricted range of rapidities $Y \,\leq\,1/\bas^4$~\cite{LEBH}.


\subsubsection{Expressing  the double diffraction in terms of the single ones} 

Generalizing the master equation of \eq{DDEQ7} to double diffraction is straightforward. Indeed, \eq{DDEQ2} and \eq{DDEQ3} can be rewritten in terms of $S^D$ as follows\footnote{KL stands for the Kovchegov--Levin equation for diffractive production~\cite{KOLE}.}:
\beq \label{DDEQ10}
(-1)^n\,C_n\,=\,\intl^{\epsilon+i \infty}_{\epsilon-i \infty}\,\frac{d \,\bg \,z^D_{01}}{2\,\pi\,i}\,e^{ - n\,\bg\, z^D_{01}}\!\!\!\!\!\underbrace{S^D\Lb z^D_{01}\Rb}_{\mbox{KL equation}}\,;~~(-1)^n\,C_n\,=\,\intl^{\epsilon+i \infty}_{\epsilon-i \infty}\,\frac{d \,\bg\, \bar{z}^D_{02}}{2\,\pi\,i}\,e^{ - n\,\bg\,\bar{z}^D_{02}}\,\!\!\!\!\!\underbrace{S^D\Lb \bar{z}^D_{02}\Rb}_{\mbox{KL equation}};
\eeq
where $z^D_{01} = z_{m1} + \Delta z \Lb z^{01}_d\Rb$ and $\bar{z}^D_{02} = z_{m2} + \Delta z \Lb z^{02}_d\Rb$.
Using these equations, we can rewrite \eq{DDSOL2} for double diffraction in the form:
\beq \label{DDEQ11}
 S^{DD}\Lb z_{m1}, z_{m2} ,z^{12}_d\Rb\,\,=\,\,\sum^\infty_{n=0} (-1)^n\,2^{-n} \,n! \intl^{\epsilon +i \infty}_{\epsilon-i \infty} \frac{d \,\bg\,z^D_{01}}{2\,\pi\,i}\intl^{\epsilon +i \infty}_{\epsilon-i \infty} \frac{d \,\bg\,\bar{z}^D_{02}}{2\,\pi\,i}e^{ n\,\bg\, \Lb  z^D_{12} \,-\, z^D_{01} \,-\, \bar{z}^D_{02}\Rb} \,S^D\Lb z^D_{01} \Rb\,S^D\Lb\bar{z}^D_{02}\Rb
\eeq
where $z^D_{12} = z_{m1} + z_{m2} + \Delta z\Lb z^{12}_d\Rb$. Neglecting the factor $(-1)^n 2^{-n}\,n!$ and summing over $n$ in \eq{DDEQ11}, we obtain:
\beq \label{DDEQ12}
  S^{DD}\Lb z^D_{12}\Rb\,=\,\intl^{\epsilon +i \infty}_{\epsilon-i \infty} \frac{d \,\bar{z}^D_{02}}{2\,\pi\,i} \,S^D\Lb z^D_{12} - \bar{z}^D_{02}\Rb\,S^D\Lb\bar{z}^D_{02}\Rb
\eeq
As discussed in the previous section, we need to perform the Borel transform for one of the $S^D$ in \eq{DDEQ12} (see \eq{DDEQ6}). Introducing
\beq \label{DDEQ13}
S^D\Lb e^{\bg\,z^D_{01}}\Rb\,\,=\,\,\intl^\infty_0\!d t \,e^{-t} {\cal B}S^D\Lb - t e^{\bg\,z^D_{01}}\Rb
\eeq
we can rewrite \eq{DDEQ12} in the form:
\beq \label{DDEQ14}
  S^{DD}\Lb z^D_{12}\Rb\,=
\,\,\intl^\infty_0 d t \,e^{-t}\, \frac{1}{\bg}\intl^{\epsilon +i \infty}_{\epsilon-i \infty} \frac{d \,\bar{z}^D_{02}}{2\,\pi\,i} \underbrace{{\cal B}S^D\Lb \ln\Lb - t/2\Rb/\bg +  z^D_{12}  - \bar{z}^D_{02}\Rb}_{\mbox{KL equation}}\,\underbrace{S^D\Lb\bar{z}^D_{02}\Rb}_{\mbox{KL equation}} \eeq
This equation shows that for double diffraction we obtain an expression very similar to \eq{DDEQ7}, with $S \to S^D$ and $z^{01}_d \to z^D_{01}$ ($\bar{z}^{02}_d \to \bar{z}^D_{02}$).

\section{Conclusions} 


This paper presents a first attempt to investigate double diffractive production in dipole-dipole scattering within high energy QCD. Our approach is based on summing large BFKL Pomeron loops, which are responsible for producing two clusters of hadrons separated by a rapidity gap. Notably, in our framework both clusters can carry large masses --- something that is not possible within the Balitsky-Kovchegov (BK) approach.

To sum the large loops, we utilize the dipole density extracted from the solution to the single diffraction equation~\cite{KOLE} (see \eq{KL}), along with the simple nuclear model given by \eq{DD41}.

We develop two model approaches to double diffractive production: (i) a one dimensional model, and (ii) QCD with a simplified leading-twist BFKL kernel. These two models clarify the main features of double diffractive production and yield analytical results across the entire kinematic region of produced masses.

Our first main result is that the single diffraction cross section $\sigma^{D}$ exhibits geometric scaling: it depends only on the single variable
$z_m + 2\sqrt{\kappa\Omega(z^{01}_d)}$ (see \eq{BK20} and \eq{BK21}). The double diffraction cross section $\sigma^{DD}$ shows the same behaviour, depending only on the combination $z_{m1} + z_{m2} + \Delta z (z^{12}_d)$, where $\Delta z(z^{12}_d) = \sqrt{2}\, z^{12}_d$ for large $z^{12}_d$ (see \eq{DDSOL7B}).
 
Our second main result is the following expression for $S^{DD}$:
\beq \label{C1}
S^{DD}(z_{m1}+z_{m2} +\Delta z (z^{12}_d)) = \Phi(z_{m1}+z_{m2} +\Delta z(z^{12}_d)) \exp\left( - \frac{ (z_{m1}+z_{m2} +\Delta z (z^{12}_d) - 8 - 2 \ln 2)^2}{4\,\kappa}\right),
\eeq
where $\Delta z (z^{12}_d) = 2 \sqrt{\kappa\Omega(z^{12}_d)}$  and $\Phi$ is a smooth function of its argument.

For the regular BFKL kernel, we find that the cross section is given by \eq{C1} with $\Delta z (z^{12}_d) = \sqrt{2} z^{12}_d$ for large values of $z_{m1}, z_{m2}$ and $z_{d}^{12}$. We also propose an equation for double diffractive production that satisfies the appropriate initial conditions.  

Crucially, we have verified each step of our approach against the one  dimensional model, which admits an exact solution, and found that all the intuition gained from this simple model carries over to our approach in QCD.
 
 The process of double diffractive production into two high-mass clusters lies beyond the reach of the BK approach, so experimental data on this process offer a genuine test of how well the summed large Pomeron loops describe it. Confronting the predicted scaling behaviour with data from high-energy colliders such as the LHC~\cite{ALICE} will therefore be an important step towards establishing whether large Pomeron loops play a significant role in high energy QCD dynamics, and in particular in the unitarization of the scattering amplitude. 

~
\section{Acknowledgements}
We thank our colleagues at Tel Aviv University and UTFSM for encouraging discussions.  This research was supported by Fondecyt (Chile) grant No. 1231829. J. G. expresses his gratitude to the Institute of Physics of PUCV (IFIS).
 

   ~
  
 \appendix
  \section{Solution to the BFKL equation}
  \label{appendix:A}
 The BFKL equation for the elastic scattering amplitude ($N$) has the form:

\beq \label{SOL1}
\frac{\partial N^{\rm BFKL} \Lb \vec{x}_0,\vec{x}_1, Y\Rb}{\partial Y}\,\,= \,\,\frac{\bas}{2
\,\pi}\,\int d^2\,x_{2} \frac{ x^2_{01}}{x^2_{02}\,x^2_{12}}\,\Big\{ N^{\rm BFKL} \Lb \vec{x}_1,\vec{x}_2, Y\Rb \,\,+\,\,N^{\rm BFKL} \Lb \vec{x}_0,\vec{x}_2, Y\Rb\,\,-\,\,N^{\rm BFKL} \Lb \vec{x}_0,\vec{x}_1, Y\Rb\Big\}
\eeq
The eigenfunctions of these equations have the following form~\cite{LIP}:
\begin{subequations}
\bea
\phi_\gamma\Lb \vec{r}, \vec{r}_P, \vec{b}\Rb &=& \Lb \frac{r^2\,r^2_P}{\Lb \vec{b} - \h\Lb \vec{r}-\vec{r}_P\Rb\Rb^2 
\,\Lb \vec{b} + \h\Lb \vec{r}-\vec{r}_P\Rb\Rb^2 }\Rb^\gamma\,\,=\,\,e^{\gamma\,\xi_{\vec{r},\vec{r}_P}}; \label{EF}\\
\mbox{with}~~ \xi_{\vec{r},\vec{r}_P} &=& \ln\Lb \frac{r^2\,r^2_P}{\Lb \vec{b} - \h\Lb \vec{r}-\vec{r}_P\Rb\Rb^2 
\,\Lb \vec{b} + \h\Lb \vec{r}-\vec{r}_P\Rb\Rb^2 }\Rb \label{XI};
\eea
\end{subequations}
In fact, it is proven in Ref.~\cite{LIP} that this eigenfunction holds for any kernel that respects conformal symmetry. The eigenvalue for the BFKL equation has the form\footnote{$\gamma$ is the anomalous dimension in these formulae. We hope that it will not be confused with the notation for scattering amplitude in the main text.}:
\beq \label{EV}
\omega\Lb \bas, \gamma\Rb \,\,=\,\,\bas \chi\Lb \gamma\Rb \,\,=\,\, \bas\Lb 2 \psi(1) - \psi\Lb \gamma\Rb  -  \psi\Lb 1 - \gamma\Rb\Rb\,\,\xrightarrow{\gamma \to \h}\,\,\underbrace{ 4 \,\ln 2\,\bas}_{\Delta_{\pom}} \,-\,\psi ^{(2)}\left(\frac{1}{2}\right)\,\bas\,\left(\gamma -\frac{1}{2}\right)^2 \,+\,\dots
\eeq
where $\psi(z)$ denotes the Euler psi-function: $\psi(z) = d \ln \Gamma\Lb z\Rb/d z$ (see \cite{RY}). In \eq{SOL1}
 we denote  by $\vec{x}_1 $ and $\vec{x}_0$ the coordinates of the quark and antiquark in the colourless dipole, respectively. In \eq{EF} and    \eq{XI} we introduce: (i) the size of the dipole: $\vec{r} = \vec{x}_1 - \vec{x}_0$, (ii) the size of the dipole in the projectile (or target) $\vec{r}_P = \vec{x}_{1,P} - \vec{x}_{0,P}$ and (iii) the impact parameter $\vec{b} = \h\Lb \vec{x}_1 +\vec{x}_0 - \vec{x}_{1,P} - \vec{x}_{0,P}\Rb$. 
 
 For  the interaction with nucleus, which we consider in this paper,   the  scattering amplitude of the dipole with a nucleus  for the exchange of the BFKL Pomeron   has the following form:
\beq \label{EIGENF1}
A^{\rm BFKL}_{d A} \Lb  \vec{r} , \vec{R}_N, \vec{b}; Y\Rb\,=\,\intl^{\epsilon + i \infty}_{\epsilon - i \infty}\!\!\!\frac{ d \gamma}{2 \pi i}\intl \!\!d^2 c\,e^{\omega\Lb \bas, \gamma\Rb\,Y}\phi_\gamma\Lb \vec{r} , \vec{R}_N, \vec{c}\Rb\,\,S_A\Lb \vec{b} - \vec{c}\Rb\eeq
where $R_N$ is the size of a nucleon, $\vec{c}$ is the impact parameter for dipole-nucleon amplitude and $\vec{b} - \vec{c}$ is the position of the nucleon with respect to the center  of the nucleus. $S_A\Lb \vec{b} - \vec{c}\Rb$ is the number of  nucleons in a nucleus.  Since $c \,\ll \,b$   we can integrate over $\vec{c}$ replacing $\vec{b} - \vec{c} $ by $\vec{b}$ and obtain  the expression:
\bea \label{EIGENF2}
A^{\rm BFKL}_{d A} \Lb  \vec{r} , \vec{R}_N, \vec{b}; Y\Rb  &=&\intl^{\epsilon + i \infty}_{\epsilon - i \infty}\!\!\!\frac{ d \gamma}{2 \pi i}\intl \!\!d^2 c \,e^{\omega\Lb \bas, \gamma\Rb\,Y}\phi_\gamma\Lb \vec{r} , \vec{R}_N, \vec{c}\Rb\,\,S_A\Lb \vec{b} - \vec{c}\Rb = \intl^{\epsilon + i \infty}_{\epsilon - i \infty}\!\!\!\frac{ d \gamma}{2 \pi i}\intl\!\! d^2 c \, e^{\omega\Lb \bas, \gamma\Rb\,Y}\phi_\gamma\Lb \vec{r} , \vec{R}_N, \vec{c}\Rb\,\,S_A\Lb \vec{b}\Rb \nn\\
&=&  \Lb e^{\frac{\omega\Lb \bas, \bg\Rb}{\bg}\,Y} \frac{r^2}{R^2_N}\Rb^{\bg} \!\!R^2_N S_A\Lb \vec{b}\Rb   =   \Lb r^2\,Q^2_s\Lb Y, \vec{R}_N, \vec{b}\Rb \Rb^{\bg}\,\eea
Therefore, in this case we can absorb all dependence on the impact parameter in the $b$  dependence of the saturation scale. In \eq{EIGENF2} we implicitly assume that $ r \leq R_N$. 
 One can see that \eq{EIGENF2} reduces to \eq{DD41} which we use to obtain the dipole densities from the dipole-nucleus scattering. In \eq{Z} the size of large dipole $R = R_A = A^{1/3}R_N$.

  \section{Elastic dipole-dipole scattering}
  \label{appendix:B}
 
 First, we consider the elastic scattering of a dipole $r$ with a nucleus, which satisfies the BK equation. The BK equation has the same form as \eq{KL1} but for $S$, which is given by \eq{DD13}. 
$S^{d}_{A}\Lb z^{01}_d\Rb$ is a series of multi-BFKL Pomeron exchanges, viz.:
\beq \label{QCD4}
S^{d}_{A}\Lb z^{01}_d\Rb\,\,=\,\,\sum_{n=0}^{\infty} (-1)^n\,C_n \,e^{n\,\bar{\gamma}\,z^{01}_d}~~~~e^{n\,\bar{\gamma}\,z^{01}_d}\,=\mbox{Green's function of $n$-BFKL Pomeron exchanges}
\eeq
Plugging this equation in \eq{QCD1} for the elastic $S$
we obtain:
\beq \label{QCD5A}
\sum_{n=0}^{\infty} (-1)^n\,n\,C_n e^{n\,\bar{\gamma}\,z^{01}_d}\,\,=\,\, - \frac{z^{01}_d}{\kappa\,\bg} \sum_{n=0}^{\infty} (-1)^n\,C_n e^{n\,\bar{\gamma}\,z^{01}_d}~~~~\mbox{or}~~~~ \langle|n|\rangle\,\,=\,\,- \frac{z^{01}_d}{\kappa \bg}\eeq
We can understand this equation if we rewrite the sum over $n$ as the integration over contour $C$ in \fig{cont}, viz.:
 \beq \label{QCD5}
 S^{d}_A\Lb z^{01}_d\Rb  =\oint_C \frac{d n}{2\,\pi\,i}  C_n \frac{\pi}{\sin\Lb \pi\,n\Rb} G^n_{\pom}\Lb z^{01}_d\Rb\eeq
Assuming that we can open the contour $C$  to the contour $C'$  we can take the integral using the method of steepest descent.  Denoting $C_n =\exp\Lb \phi(n)\Rb$ we obtain the equation for the saddle point $n_{SP}$:
\beq \label{QCD6}
\phi'_n \Lb n_{SP}\Rb=- \bar{\gamma} \,z^{01}_d\eeq
From \eq{QCD4} $n_{SP} = - \frac{z}{\bg\,\kappa}$ and \eq{QCD6}  gives   $\phi\Lb n\Rb= \bar{\gamma}^2 \kappa \frac{n^2}{2}$.

Therefore, \eq{RHONF} for  $  \rho_n\Lb Y_{01} - Y_0, \vec{r},\vec{b};\{ \vec{r}_i,\vec{b}_i\}\Rb$ has the form:
\beq \label{RHONFD} 
   \rho_n\Lb Y_{01} - Y_0,  \vec{r},\vec{b};\{ \vec{r}_i,\vec{b}_i\}\Rb\,\,=\,\,e^{ \bar{\gamma}^2 \kappa \frac{n^2}{2}}  \prod_{i=1}^n G_{\pom} \Lb z^{01}_{i ,d}\Rb~~~~\mbox{with}~~~~ z^{01}_{i ,d} \,= \,\kappa\,\bas\,\Lb Y_{01}- Y_0\Rb \,+\,\xi_{\vec{r}_i, \vec{r}}
   \eeq
Calculating $S^d_d\Lb z^{12}_d\Rb$ using \eq{MPSI}, \eq{UT} and  \eq{RHONFD} we obtain
\beq \label{QCD7}
S^d_d\Lb z^{12}_d\Rb\,\,=\,\,\int^\infty_0 d t e^{-t}\sum_{n=0}^{\infty} (-1)^n\,C^2_n\,t^n\, e^{n\,\bar{\gamma}\, z^{12}_d}=\int^\infty_0 d t e^{-t}\sum_{n=0}^{\infty} (-1)^n\,e^{ 2 \phi\Lb n\Rb}\,t^n\, e^{n\,\bar{\gamma}\, z^{12}_d}
\eeq
In this equation we use $ z^{01}_{i ,d} \,= \,\kappa\,\bas\,\Lb Y_{01}- Y_0\Rb \,+\,\xi_{\vec{r}_i, \vec{r}} $, $z^{02}_{i ,d} \,= \,\kappa\,\bas\,\Lb Y_{0}- Y_{02}\Rb \,+\,\xi_{\vec{r}'_i, \vec{r}'}$ and $z^{12}_d$ from \eq{Z}. We use \eq{UT} to make the integrals over $r_i$ and $b_i$.

Taking integral over $t$ we have:
\beq \label{QCD8}
S^d_d\Lb z^{12}_d\Rb\,\,=
\,\,\sum_{n=0}^{\infty} (-1)^n\,C^2_n\,\Gamma\Lb n+1\Rb e^{n\,\bar{\gamma}\, z^{12}_d}=\sum_{n=0}^{\infty} (-1)^n\,e^{ 2 \phi\Lb n\Rb}\,n!\, e^{n\,\bar{\gamma}\ z^{12}_d}
\eeq

Rewriting this integral as a contour integral (see \eq{QCD5}), 
and taking the integral over $n$ using contour $C''$ in \fig{cont}, 
we can write \eq{QCD8} in the form:
\beq \label{QCD9}
S^d_d\Lb z^{12}_d\Rb\,\,=
\,\,\oint_{C''}\frac{dn}{2\,\pi\,i}\,C^2_{n}\,\frac{\pi\,\Gamma\Lb 1 +n\Rb}{\sin\Lb \pi\,n\Rb}\, e^{n\,\bar{\gamma}\, z^{12}_d}\,\,=\,\,
\sum_{k=1}^{\infty} \,\frac{d}{d k}\left[\frac{1}{\Gamma\Lb k \Rb}\,e^{2 \phi\Lb -k\Rb\,-\,k\,\bar{\gamma}\, z^{12}_d}\right]\,\,=\,\,\sum_{k=1}^{\infty} \frac{\Phi(k)}{\Gamma\Lb k\Rb}\,e^{\bar{\gamma}^2 \kappa k^2\,-\,k\bar{\gamma}z^{12}_d}
\eeq

where $\Phi\Lb  k\Rb= \psi ^{(0)}(k)- 2\,\bg^2 \kappa k - \bar{\gamma}\, z^{12}_d$ is the residue of the double pole. In \eq{QCD9} we use that $\Gamma( n \to - k) = \frac{(-1)^k}{k!\,(n+k)}$ (see Ref.~\cite{RY}, formula {\bf 8.310}), where $k$ is a positive integer.
In \eq{QCD9} we also used that $\phi\Lb -k\Rb=\phi\Lb k \Rb$. Taking the sum in this equation by the method of steepest descent, we find that
\beq \label{QCD10}
S^{ d}_{ d}\Lb  z^{12}_d\Rb\,\,=\,\,\frac{\Phi\Lb  z^{12}_d\Rb}{\Gamma\Lb \frac{ z^{12}_d}{2\,\bg\,\kappa}\Rb} \exp\Lb -\, \frac{ \Lb z^{12}_d\Rb^2}{4\,\kappa}\Rb
\eeq
where $\Phi$ is a smooth function of $ z^{12}_d$.


\section{Borel transform of the elastic S-matrix} 
\label{appendix:C}
For a function $A(z)=\sum_{n=0}^\infty a_n z^n$, the Borel representation reads \cite{BOREL}
\beq\label{QCD11}
A(z)\,=\,\sum_{n=0}^\infty a_n\,z^n\,=\,\int\limits_0^\infty dt\,e^{-t}\,\mathcal{B}A(tz),
\qquad
\mathcal{B}A(tz)\,=\,\sum_{n=0}^\infty a_n\,\frac{(tz)^n}{n!},
\eeq
which follows from $\int_0^\infty dt\, e^{-t}\,t^n = n!$.

We apply this to the elastic $S$-matrix, taking $G_{\pom}(z)=e^{\bar\gamma z}$ and
$C_n=e^{\bar\gamma^2\kappa\,n^2/2}$:
\beq\label{QCD12}
S(z)\,=\,\sum_{n=0}^\infty(-1)^n\,C_n\,G_{\pom}^n(z)\,=\,\sum_{n=0}^\infty C_n\left(-G_{\pom}(z)\right)^n
\,=\,\int\limits_0^\infty dt\,e^{-t}\,\mathcal{B}S\!\left(-t\,e^{\bar\gamma z}\right),
\eeq
where the Borel image is
\beq\label{QCD13}
\mathcal{B}S\!\left(-t\,e^{\bar\gamma z}\right)\,=\,\sum_{n=0}^\infty C_n\,\frac{\left(-t\,e^{\bar\gamma z}\right)^n}{n!}
\,=\,\sum_{n=0}^\infty \frac{(-1)^n}{n!}\,t^n\,e^{\bar\gamma^2\kappa\,\frac{n^2}{2}}\,e^{n\bar\gamma z}.
\eeq

To perform the sum over $n$, we linearize the quadratic exponent with the Gaussian (Hubbard--Stratonovich) identity, valid for $a>0$:
\beq\label{QCD14}
e^{a\,n^2}\,=\,\frac{1}{\sqrt{4\pi a}}\intl_{-\infty}^{\infty}\!d\lambda\;e^{-\frac{\lambda^2}{4a}\,+\,n\lambda}.
\eeq
Setting $a=\bar\gamma^2\kappa/2$ and summing the resulting exponential series in $n$, we obtain
\bea\label{QCD15}
\mathcal{B}S\!\left(-t\,e^{\bar\gamma z}\right)
&=&\frac{1}{\sqrt{2\pi\,\bar\gamma^2\kappa}}\intl_{-\infty}^{\infty}\!d\lambda\,
\exp\!\left(-\frac{1}{\bar\gamma^2\kappa}\Lb\frac{\lambda^2}{2}\,+\,\eta(z)\,e^{\lambda}\Rb\right),
\qquad \eta(z)\,=\,\bar\gamma^2\kappa\,t\,e^{\bar\gamma z}.
\eea

The integral is evaluated by the saddle-point method, which is justified by the large factor $1/(\bar\gamma^2\kappa)$ in the exponent. The saddle-point condition is
\beq\label{QCD16}
\lambda_{SP}\,+\,\eta(z)\,e^{\lambda_{SP}}\,=\,0,
\eeq
with the solution
\beq\label{QCD17}
\lambda_{SP}\,=\,-W\!\left(\eta(z)\right),
\eeq
where $W$ is the Lambert $W$ function, defined by $W(\eta)\,e^{W(\eta)}=\eta$. For large $\eta$ it behaves as
$W(\eta)\simeq\ln\eta-\ln\!\left(\ln\eta\right)$. Since $\ln\eta=\bar\gamma z+\ln(\bar\gamma^2\kappa t)$, at large $z$ we have
$\lambda_{SP}(z)\simeq-\bar\gamma z$, up to terms that depend only logarithmically on $z$ and $t$.

Evaluating the Gaussian integral around the saddle point, and using $\eta e^{\lambda_{SP}}=-\lambda_{SP}$ together with $\partial^2_\lambda\!\left(\lambda^2/2+\eta e^\lambda\right)\big|_{SP}=1-\lambda_{SP}$, gives
\bea\label{QCD18}
\mathcal{B}S\!\left(-t\,e^{\bar\gamma z}\right)
&\simeq&\frac{1}{\sqrt{1-\lambda_{SP}}}\,\exp\!\Lb-\frac{\lambda_{SP}^2-2\lambda_{SP}}{2\,\bar\gamma^2\kappa}\Rb
\,\simeq\,\Psi(z)\,\exp\!\Lb-\frac{z^2}{2\kappa}\Rb,
\eea
where $\Psi(z)$ is a smooth function of $z$. Substituting $\lambda_{SP}\simeq-\bar\gamma z$ shows that it collects the prefactor and the term linear in $z$ of the exponent, $\Psi(z)\simeq(1+\bar\gamma z)^{-1/2}\,e^{-z/(\bar\gamma\kappa)}$. Thus the Gaussian fall-off $e^{-z^2/2\kappa}$ dominates the large-$z$ behavior of $S$.


\end{document}